\documentclass[journal]{IEEEtran}
\usepackage[cmex10]{amsmath}
\usepackage{graphicx,amssymb,amsthm,comment,bm,cite,url}
\usepackage{color}
\usepackage{boldline}
\usepackage{algorithm,algorithmic}
\usepackage{multirow}
\allowdisplaybreaks

\newcommand{\refeq}[1]{Eq. (\ref{eq:#1})}

\newcommand{\refsec}[1]{Section \ref{sec:#1}}
\newcommand{\refsubsec}[1]{Subsection \ref{subsec:#1}}

\newcommand{\reffig}[1]{Fig. \ref{fig:#1}}

\def\Vec#1{\boldsymbol{\mathbf{#1}}}
\def\Argmax{\mathop{\rm argmax}}

\def\thline{\noalign{\hrule height 1.2pt}}

\def\Xsrc{\Vec{X}^{(\mathsf{s})}}
\def\Xtrg{\Vec{X}^{(\mathsf{t})}}
\def\xsrc{\Vec{x}^{(\mathsf{s})}}
\def\xtrg{\Vec{x}^{(\mathsf{t})}}

\def\Nsrc{N_{\mathsf{s}}}
\def\Ntrg{N_{\mathsf{t}}}

\ifCLASSINFOpdf
\else
\fi

\begin{document}
%
\title{
ConvS2S-VC: Fully Convolutional Sequence-to-Sequence Voice Conversion
}
%
%
%

\author{Hirokazu~Kameoka,
Kou Tanaka,
Damian Kwa\'sny,
Takuhiro Kaneko,
and Nobukatsu Hojo%
\thanks{H. Kameoka, K. Tanaka, Damian Kwa\'sny, 
T. Kaneko and N. Hojo are with NTT Communication Science Laboratories, Nippon Telegraph and Telephone Corporation, Atsugi, Kanagawa, 243-0198 Japan (e-mail: hirokazu.kameoka.uh@hco.ntt.co.jp).}
\thanks{This work was supported by JSPS KAKENHI 17H01763
and JST CREST Grant Number JPMJCR19A3, Japan. 
}}

%
%

\markboth{}%
{}
%



\maketitle

\begin{abstract}
This paper proposes a voice conversion (VC) method using
sequence-to-sequence (seq2seq or S2S) learning, 
which flexibly converts not only the
voice characteristics but also the pitch contour and duration of input speech.
The proposed method, called ConvS2S-VC,  
has three key features. 
First, it uses a model with a fully convolutional architecture. 
This is particularly advantageous in that it is suitable for parallel computations using GPUs.
It is also beneficial since it enables effective normalization techniques such as batch normalization to be used for all the hidden layers in the networks. 
Second, it achieves 
many-to-many
conversion 
by simultaneously learning mappings among multiple
speakers
using only a single model
instead of separately learning mappings between each
speaker
pair using a different model.
This enables the model to fully utilize available training data collected from multiple
speakers
by capturing common latent features that can be shared across different 
speakers.
Owing to this structure, 
our model works reasonably well even without source 
speaker
information, 
thus making it able to handle any-to-many conversion tasks.
Third, we introduce a mechanism, 
called
the conditional batch normalization
that switches batch normalization layers in accordance with the target speaker.
This particular mechanism has been found to be extremely effective for our 
many-to-many conversion model.
We conducted speaker identity conversion experiments
and found that ConvS2S-VC
obtained higher sound quality and speaker similarity than baseline methods.
We also found from audio examples that 
it could perform well in 
various tasks including emotional expression conversion, 
electrolaryngeal speech enhancement, and English accent conversion.
\end{abstract}


%
\IEEEpeerreviewmaketitle

\section{Introduction}
\label{sec:intro}

Voice conversion (VC) is a technique for converting 
para/non-linguistic information contained in a given utterance 
such as the perceived identity of a speaker
while preserving linguistic information.
Potential applications of this technique include
speaker-identity modification \cite{Kain1998short}, 
speaking aids \cite{Kain2007short,Nakamura2012short}, 
speech enhancement \cite{Inanoglu2009short,Turk2010short,Toda2012short}, 
and accent conversion \cite{Felps2009short}.

Many conventional VC methods 
are designed to use parallel utterances of source and target speech to 
train acoustic models for feature mapping. 
A typical pipeline of the training process  
consists of extracting acoustic features from source and target utterances,
performing dynamic time warping (DTW) to obtain time-aligned parallel data,
and training an acoustic model that maps the source features to the target features
frame-by-frame.
Examples of the acoustic model include Gaussian mixture models (GMM) \cite{Stylianou1998short,Toda2007short,Helander2010short} and deep neural networks (DNNs) 
\cite{Desai2010short,Mohammadi2014short,YSaito2017bshort,Sun2015short,Kaneko2017cshort}. 
Some attempts have also been made to 
develop methods that 
require no parallel utterances, transcriptions, or time alignment procedures.
Recently, deep generative models such as 
variational autoencoders (VAEs), 
cycle-consistent generative adversarial networks (CycleGAN),
and star generative adversarial networks (StarGAN) 
have been used with notable success for non-parallel VC tasks \cite{Hsu2016short, Hsu2017short, Kameoka2019IEEETransshort_ACVAE-VC, Kaneko2017dshort, Kameoka2018SLTshort_StarGAN-VC}.

One limitation of conventional methods 
including those mentioned above
is that they are focused mainly on learning to convert only the local spectral features 
and less on converting prosodic features 
such as the fundamental frequency ($F_0$) contour, duration, and rhythm of the input speech.
This is because 
the acoustic models in
these methods are designed to describe mappings between local features only.
This prevents a model from discovering
word-level or sentence-level 
suprasegmental conversion rules.
In most methods, 
the entire $F_0$ contour is simply adjusted using a linear 
transformation in the logarithmic domain 
while the duration and rhythm are usually kept unchanged. 
However, since these features play as important a role as local spectral features
in characterizing speaker identities and speaking styles, 
it would be desirable if these features could also be converted more flexibly.
To overcome this limitation, 
we need a model that 
can learn to convert entire feature sequences
by capturing and utilizing 
long-term dependencies in source and target speech.
To this end, we adopt a sequence-to-sequence (seq2seq or S2S) learning approach.

The S2S learning approach  
offers a general and powerful framework for transforming 
one sequence into another variable length sequence \cite{Sutskever2014short,Chorowski2015NIPSshort}.
This is made possible by using encoder and decoder networks,
where the encoder encodes an input sequence to an internal representation 
whereas the decoder generates an output sequence in accordance with the internal representation. 
The original S2S model employs recurrent neural networks (RNNs) to model the
encoder and decoder networks, where
common choices for the RNN architectures involve
long short-term memory (LSTM) networks and gated recurrent units (GRU).
This approach has  
attracted a lot of attention in recent years after being
introduced and applied with notable success in various tasks such 
as machine translation,
automatic speech recognition (ASR) \cite{Chorowski2015NIPSshort} and
text-to-speech (TTS) \cite{Wang2017short,Arik2017ashort,Arik2017bshort,Sotelo2017short,Tachibana2018short,Ping2018ICLRshort,Shen2018ICASSPshort}. 

The original S2S model 
suffers from the constraint that all input sequences are forced to be encoded into a fixed length internal vector.
This limits the ability of the model
especially when it comes to long input sequences, such as long sentences in text translation problems. 
To overcome this limitation, 
a mechanism called ``attention'' \cite{Luong2015short} has been introduced, 
which enables the network to learn where to pay attention in the input sequence for each item in the output sequence.

While RNNs are a natural
choice for modeling long sequential data, 
recent work has shown that convolutional neural networks (CNNs) 
with gating mechanisms also have excellent potential
for capturing long-term dependencies \cite{Dauphin2017short,vandenOord2016short}. 
In addition, they are suitable for parallel computations using GPUs unlike RNNs.
To exploit this advantage of CNNs, 
an S2S model was recently proposed
that adopts 
a fully convolutional architecture 
\cite{Gehring2017ICMLshort}. 
With this model, 
the decoder is designed using causal convolutions 
so that it enables the model
to generate an output sequence autoregressively. 
This model with an attention mechanism is called the ConvS2S model 
and has already been applied successfully  
to machine translation \cite{Gehring2017ICMLshort} and TTS \cite{Tachibana2018short,Ping2018ICLRshort}.
Inspired by its success in these tasks, 
we propose a VC method based on the ConvS2S model,
which we call ConvS2S-VC, along with an architecture 
tailored for use with VC. 

In a wide sense,
VC is a task of converting the domain of speech.
Here, the types of domain include 
speaker identities, emotional expressions, speaking styles, and accents, 
but for concreteness, 
we will restrict our attention to 
speaker identity conversion tasks
in the following.
When we are interested in converting speech among multiple speakers,
one naive way of applying the S2S model 
is to prepare and train a model for each speaker pair. 
However, 
this can be inefficient
since
the model for one pair of speakers 
fails to use the training data of the other speakers for training,
even though there must be a common set of latent features that can be shared across different 
speakers, especially when the languages are the same.
To fully utilize available training data collected from multiple speakers, 
we further propose an extension of 
the ConvS2S model that allows for ``many-to-many'' VC, 
which can learn 
mappings among multiple speakers using only a single model.

One important advantage of using fully convolutional networks
is that it enables the use of batch normalization in all the hidden layers. 
This is practically beneficial since 
batch normalization is known to be significantly effective in 
not only accelerating training but also
improving the generalization ability of the resulting models.
Indeed,
as described later,
it also positively affected our pairwise model. 
However, 
as for the many-to-many model, 
the distributions of the layer inputs can change depending on the source and target 
speakers,
which may affect model training. 
To stabilize layer input distributions,
we introduce a mechanism, called the conditional batch normalization,
that switches batch normalization layers in accordance with the source and target 
speakers.
This particular mechanism was experimentally found to work very well.

\section{Related work}

Note that 
some attempts have recently been made to apply S2S models to VC problems,
including the ones we proposed previously \cite{Kameoka2018arXivshort_ConvS2S-VC, Tanaka2019short}. 
Although most S2S models typically require sufficiently large parallel corpora
for training, collecting a sufficient number of parallel utterances
is not always feasible.
Thus, particularly in VC tasks,
one challenge is how best to train S2S models using 
a limited amount of training data.

One idea involves using text labels as auxiliary information
for model training, assuming they are readily available. 
For example, 
Miyoshi et al. proposed combining 
acoustic models for ASR and TTS 
with an S2S model \cite{Miyoshi2017short},
where 
an S2S model is used to convert
the context posterior probability sequence produced by the ASR model
and 
the TTS model is finally used to generate 
a target speech feature sequence. 
Zhang et al. also proposed an S2S model-based VC method 
guided by an ASR system, which augments inputs with bottleneck features obtained from a pretrained ASR system
\cite{Zhang2018short}.
Subsequently, Zhang et al. proposed a shared model for
TTS and VC tasks, which enables joint training
of the TTS and VC functions \cite{Zhang2019short}.
Recently, Biadsy et al. proposed an end-to-end VC system called Parrotron,
which is designed to train the encoder and decoder along with
an ASR model on the basis of a multitask learning strategy \cite{Biadsy2019short}.
Our method differs from these methods in that our model 
does not rely on ASR or TTS models and requires no text annotations for model training.
Instead, we introduce several techniques to stabilize 
training and test prediction. 

Haque et al. proposed a method that enables many-to-many VC
similar to ours \cite{Haque2018short}.
As detailed in \refsubsec{many2one},
our many-to-many model differs in that 
it does not necessarily require source speaker information for the encoder, 
thus enabling it to also handle {\it any}-to-many VC tasks. 

In addition,
our method differs from all the methods mentioned above in that it adopts a fully convolutional model, which can be potentially advantageous in several ways, as already mentioned.

\section{ConvS2S-VC}
\label{sec:ConvS2S-VC}

\begin{figure}[t!]
\centering
\begin{minipage}[t]{.98\linewidth}
  \centerline{\includegraphics[width=.98\linewidth]{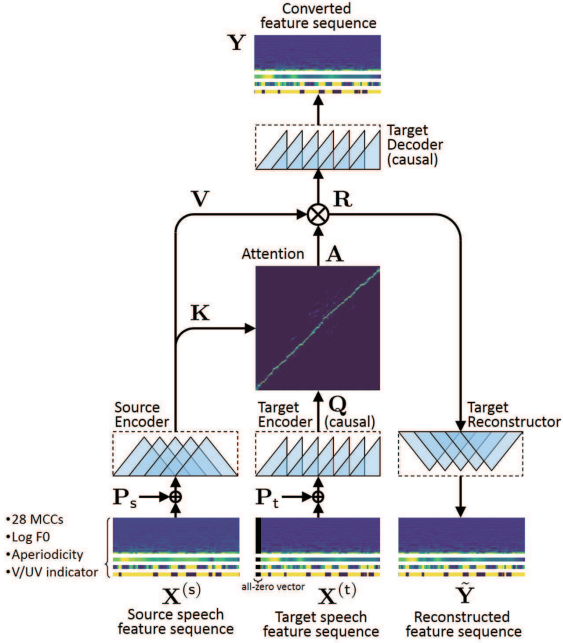}}
  \vspace{-1ex}
  \caption{Overall structure of the pairwise ConvS2S model.}
\label{fig:ConvS2S}
\end{minipage}
\vspace{-1ex}
\end{figure}

In this section, 
we start by describing a pairwise one-to-one conversion model and then present 
its multi-speaker extension that enables many-to-many VC.
The overall architecture of the pairwise conversion model is illustrated in \reffig{ConvS2S}.

\subsection{Feature extraction and normalization}
\label{subsec:feature}
First, we define acoustic features to be converted.
Although one interesting option would be to consider 
directly converting time-domain signals, 
given the recent significant advances in high-quality neural vocoder systems \cite{vandenOord2016short,Tamamori2017short,Kalchbrenner2018short,Mehri2016short,Jin2018short,vandenOord2017short,Ping2019short,Prenger2018short,Kim2018short,Wang2018short,Tanaka2018short}, 
we find it reasonable to consider converting 
acoustic features such as the 
mel-cepstral coefficients (MCCs) \cite{Fukada1992short} and log $F_0$,
since we would expect to generate high-fidelity signals 
by using a neural vocoder if we could obtain a sufficient set of acoustic features.  
In such systems, the model size for the convertor can be made small enough to 
enable the system to work well even when a limited amount of training data is available.
Hence, in this paper we choose to use the MCCs, log $F_0$, aperiodicity, and voiced/unvoiced indicator 
of speech as acoustic features as detailed below.

We first use the WORLD analyzer \cite{Morise2016short} to extract
the spectral envelope, the log $F_0$,  
the coded aperiodicity, and 
the voiced/unvoiced indicator 
within each time frame of a speech utterance, 
then compute $I$
MCCs from the extracted spectral envelope, 
and finally construct an acoustic feature vector 
by stacking
the MCCs,
the log $F_0$, 
the coded aperiodicity, 
and 
the voiced/unvoiced indicator.
Thus, each acoustic feature vector consists of $I+3$ elements.
Here, the log $F_0$ contour is assumed to be filled with smoothly
interpolated values in unvoiced segments.
At training time, 
we normalize 
each element $x_{i,n}$ $(i=1,\ldots,I)$ of the MCCs
and the log $F_0$ $x_{I+1,n}$ at frame $n$ 
to
$x_{i,n} \leftarrow (x_{i,n} - \mu_i)/\sigma_i$
where $i$, $\mu_i$ and $\sigma_i$
denote the feature index,
the mean, and the standard deviation of the $i$-th feature 
within all the voiced segments of the training samples of the same speaker.

To accelerate and stabilize training and inference,
we have found it useful to use a similar trick introduced by Wang et al. \cite{Wang2017}.
Specifically, we divide the acoustic feature sequence obtained above
into non-overlapping segments of equal length $r$
and use the stack of the acoustic feature vectors 
in each segment as a new feature vector
so that 
the new feature sequence becomes $r$ times shorter than the original feature sequence. 
Furthermore, we add the sinusoidal position encodings \cite{Vaswani2017NIPSshort} to the reshaped version of 
the feature sequence before feeding it into the model. 

\vspace{-2ex}
\subsection{Model}
\label{subsec:model}
We hereafter use
$\Xsrc=[\xsrc_1,\ldots,\xsrc_{\Nsrc}]\in \mathbb{R}^{D\times \Nsrc}$ and 
$\Xtrg=[\xtrg_1,\ldots,\xtrg_{\Ntrg}]\in \mathbb{R}^{D\times \Ntrg}$
to denote the source and target speech feature sequences of non-aligned parallel utterances,
where $\Nsrc$ and $\Ntrg$ denote the lengths of the two sequences and 
$D$ denotes the feature dimension. 
We consider an S2S model
that aims to map 
$\Xsrc$ to $\Xtrg$.
Our pairwise conversion model is inspired by and built upon the models presented
by Vaswani et al. \cite{Vaswani2017NIPSshort} and Tachibana et al. \cite{Tachibana2018short}, 
with the difference being that it involves an additional network, 
called a target reconstructor. 
This network plays an important role in ensuring that 
the encoders 
preserve contextual information about the source and target speech, as explained below.
Our model thus consists of four networks: 
source and target encoders, a target decoder, and 
a target reconstructor.

As with many S2S models, our model
has an encoder-decoder structure (\reffig{ConvS2S}).
The source and target encoders are expected to extract 
contextual information from source and target speech.
Given the contextual vector sequence pair produced by the encoders, 
we can compute a contextual similarity matrix between the source and target speech, 
which can be used to warp the time-axis of the source speech.
We can then generate the feature sequence of the target speech by letting the
target decoder transform each element of
the time-warped version of the contextual vector sequence of the source speech.
This idea can be formulated as follows.

The source encoder takes
$\Xsrc$
as the input and produces two internal vector sequences 
$\Vec{K}, \Vec{V}\in\mathbb{R}^{D'\times \Nsrc}$ 
and the target encoder takes 
$\Xtrg$
as the input and produces an internal vector sequence
$\Vec{Q}\in\mathbb{R}^{D'\times \Ntrg}$:
\begin{align}
[\Vec{K}; \Vec{V}] &= \mathsf{SrcEnc}(\Xsrc),\\
\Vec{Q} &= \mathsf{TrgEnc}(\Xtrg),
\end{align}
where 
$[;]$ denotes vertical concatenation of matrices (or vectors) 
with compatible sizes and
$D'$ denotes the dimension of the internal vectors.
$\Vec{K}$, $\Vec{V}$ and $\Vec{Q}$ 
can be metaphorically interpreted as the queries and the key-value pairs
in a hash table.
By using the query and key pair, 
we can define an attention matrix $\Vec{A}\in\mathbb{R}^{\Nsrc\times \Ntrg}$ as 
\begin{align}
\Vec{A} = \mathsf{softmax}
\big(
{\Vec{K}^{\mathsf T}\Vec{Q}}/{\sqrt{D'}}
\big),
\end{align}
where $\mathsf{softmax}$ denotes a softmax operation performed on the first axis.
$\Vec{A}$ can be seen as a similarity matrix, where the $(n,m)$-th element 
indicates the similarity between the $n$-th and $m$-th frames of source and target speech.
The peak trajectory of $\Vec{A}$ can therefore be interpreted as a time-warping function that associates 
the frames of the source speech with those of the target speech.
The time-warped version of the value vector sequence $\Vec{V}$ is thus given as 
\begin{align}
\Vec{R} = \Vec{VA},
\end{align}
which will be passed to the target decoder to generate an output sequence:
\begin{align}
\Vec{Y} = \mathsf{TrgDec}(\Vec{R}).
\end{align}

Since the target speech feature sequence 
$\Xtrg$ is of course not accessible at test time, 
we want to use a feature vector that the target decoder has generated 
as the input to the target encoder for the next time step so that 
feature vectors can be generated one-by-one recursively.
To enable the model to behave in this way, first, we must ensure 
that 
the target encoder and decoder must not use future information when producing 
an output vector at each time step. 
This can be ensured by simply 
constraining the convolution layers in the target encoder and decoder to be causal.
Note that causal convolution can be easily implemented by padding the input by 
$\delta(\kappa-1)$ elements
on both the left and right sides with zero vectors and removing $\delta(\kappa-1)$ elements from the end of the convolution output, where $\kappa$ is the kernel size and $\delta$ is the dilation factor.
Second, the output sequence $\Vec{Y}$ must correspond to 
a time-shifted version of $\Xtrg$ so that at each time step the decoder 
will be able to predict the target speech feature vector that is likely 
to be generated at the next time step.
To this end, we include an $L_1$ loss
\begin{align}
\mathcal{L}_{\mathsf{dec}} = 
{\textstyle \frac{1}{\Ntrg}}
\| \Vec{Y}_{:,1:\Ntrg-1} - \Xtrg_{:,2:\Ntrg} \|_1,
\label{eq:decloss_pairwise}
\end{align}
in the training loss to be minimized, 
where we have used the colon operator $:$ to specify the range of indices
of the elements in a matrix or a vector we wish to extract.
For ease of notation, we use $:$ itself to represent 
all elements along an axis.
For example,
$\Xtrg_{:,2:\Ntrg}$ denotes a submatrix consisting of
the elements in all the rows and columns $2,\ldots,\Ntrg$
of $\Xtrg$.
Third, the first column of $\Xtrg$ must correspond to an initial vector 
with which the recursion is assumed to start. 
We thus assume that it is always set at an all-zero vector.

The source and target encoders are free to ignore 
the information contained in
the feature vector inputs when finding 
a time alignment between source and target speech. 
One natural way to ensure that
$\Vec{K}$, $\Vec{V}$, and $\Vec{Q}$ 
contain necessary information for finding an appropriate time alignment 
is to assist 
$\Vec{K}$, $\Vec{V}$, and $\Vec{Q}$ to preserve sufficient information 
for reconstructing the input feature sequence.
To this end, 
we introduce a target reconstructor that aims to reconstruct
the feature sequence of target speech $\Xtrg$
from $\Vec{K}$, $\Vec{V}$, and $\Vec{Q}$:
\begin{align}
\widetilde{\Vec{Y}} &=
\mathsf{TrgRec}(\Vec{R}),
\end{align}
and include a reconstruction loss
\begin{align}
\mathcal{L}_{\mathsf{rec}} = 
{\textstyle \frac{1}{M}}
\|
\widetilde{\Vec{Y}}
-
\Xtrg
\|_1,
\label{eq:recloss_pairwise}
\end{align}
in the training loss to be minimized. 
This idea was introduced in our previous work \cite{Tanaka2019short}.
We call \refeq{recloss_pairwise} 
the context preservation loss. 
Although the reconstructor and the decoder may appear to have similar roles,
the difference is that 
the reconstructor is only responsible for making each column of $\Vec{R}$ contain 
sufficient information about the current value of the target feature sequence 
so that the decoder can concentrate on
predicting the future value using that information.

As detailed in \refsubsec{netarch}, 
all the networks are designed using 
fully convolutional architectures using gated linear units (GLUs) \cite{Dauphin2017short} with residual connections. 
The output of the GLU block used in the present model is defined as $\mathsf{GLU}(\Vec{X})=\mathsf{B}_1(\mathsf{L}_1(\Vec{X})) \odot \mathsf{sigmoid}(\mathsf{B}_2(\mathsf{L}_2(\Vec{X})))$ 
where $\Vec{X}$ is the layer input, $\mathsf{L}_1$ and $\mathsf{L}_2$ are dilated convolution layers, 
$\mathsf{B}_1$ and $\mathsf{B}_2$ are batch normalization layers, and $\mathsf{sigmoid}$ is a sigmoid gate function.
Similar to LSTMs,
GLUs can reduce 
the vanishing gradient problem for deep architectures
by providing a linear path for the gradients 
while retaining non-linear capabilities.

\begin{figure}[t!]
\centering
\begin{minipage}{.49\linewidth}
  \centerline{\includegraphics[width=.99\linewidth]{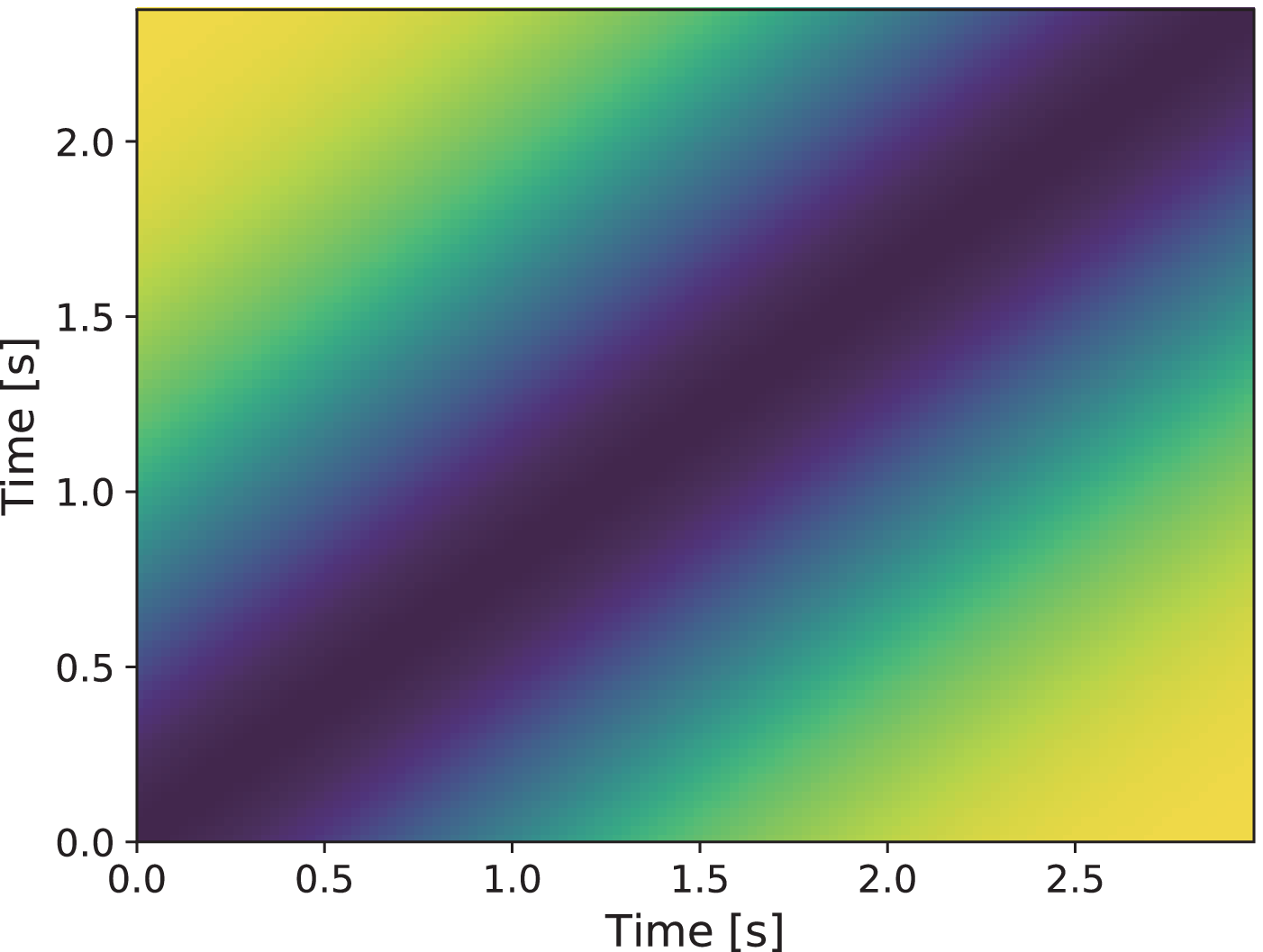}}
\end{minipage}
\begin{minipage}{.49\linewidth}
  \centerline{\includegraphics[width=.99\linewidth]{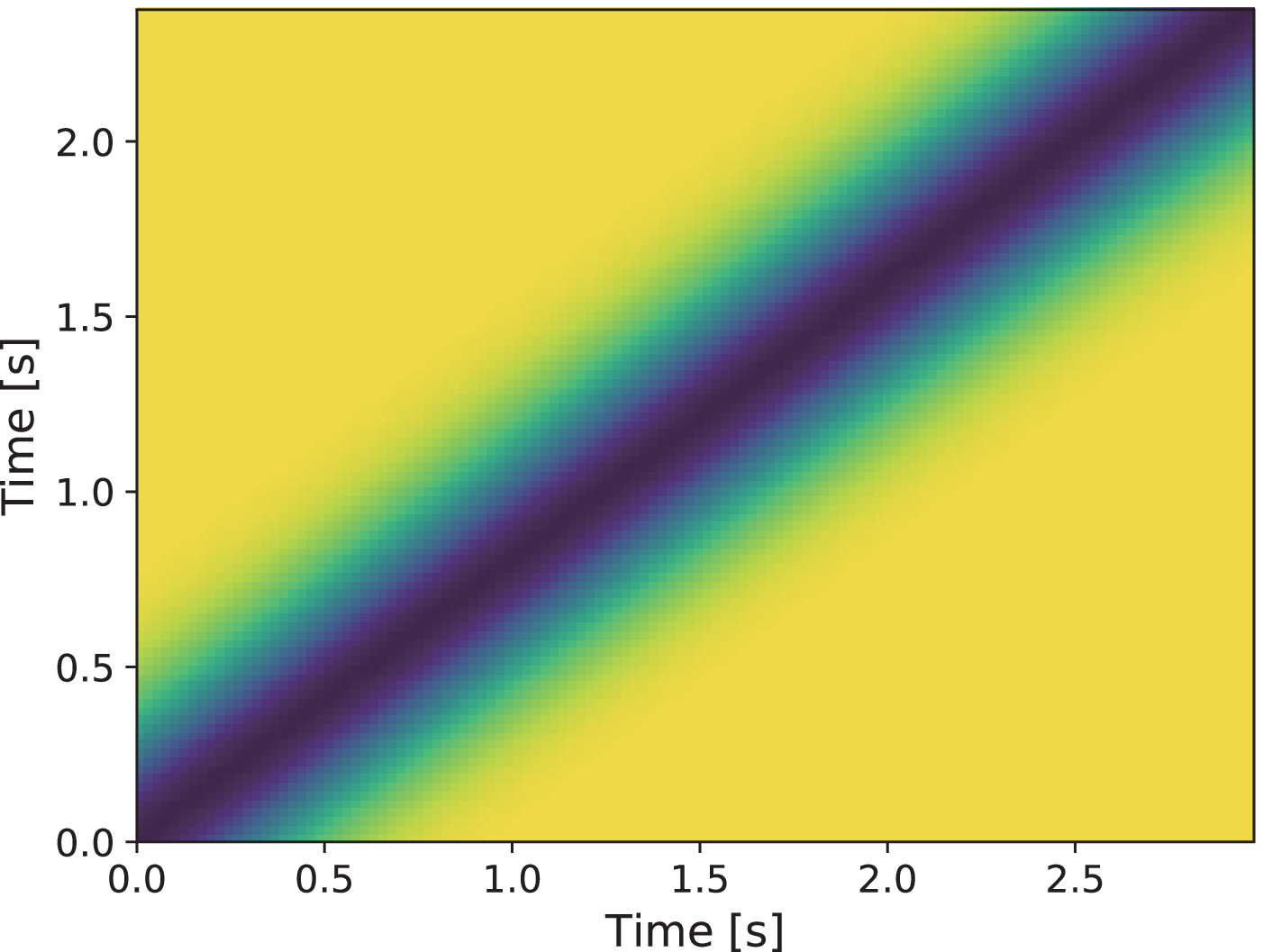}}
\end{minipage}
\vspace{-1ex}
\caption{Plots of 
$\Vec{W}_{\Nsrc\times \Ntrg}(0.3)$ (left) and 
$\Vec{W}_{\Nsrc\times \Ntrg}(0.1)$ (right)
where 
the lengths of the source and target speech 
are 2.4[s] and 3.0[s], respectively.}
\label{fig:guided_attention}
\vspace{-1ex}
\end{figure}

\subsection{Constraints on Attention Matrix}
\label{subsec:AL}

It would be natural to assume that 
the time alignment between parallel utterances is usually monotonic and nearly linear. 
This implies that the diagonal region in the attention matrix $\Vec{A}$ should always be dominant.
We expect that imposing such restrictions on $\Vec{A}$ can significantly reduce the training effort 
since the search space for $\Vec{A}$ can be greatly reduced.
To penalize $\Vec{A}$ for not having a diagonally dominant structure, 
we introduce a diagonal attention loss (DAL) \cite{Tachibana2018short}:
\begin{align}
\mathcal{L}_{\mathsf{dal}} = 
{\textstyle \frac{1}{\Nsrc\Ntrg}}
\|\Vec{W}_{\Nsrc\times \Ntrg}(\nu)\odot \Vec{A}\|_1,
\end{align}
where $\odot$ is the elementwise product and 
$\Vec{W}_{\Nsrc\times \Ntrg}(\nu)\in\mathbb{R}^{\Nsrc\times \Ntrg}$ 
is a non-negative weight matrix 
whose $(n,m)$-th element $w_{n,m}$ is defined as 
$w_{n,m} = 1- e^{-(n/\Nsrc -m/\Ntrg)^2/2\nu^2}$.
\reffig{guided_attention} shows plots of 
$\Vec{W}_{\Nsrc\times \Ntrg}(\nu)$.


Each time point of the target feature sequence must correspond to 
only one or at most a few time points of the source feature sequence.
This implies that two different columns in ${\bf A}$ must be as 
orthogonal as possible.
Although 
the DAL with a sufficiently small $\nu$ value can induce 
orthogonality, it may also lead to undesirable situations
where 
the time alignment between the two sequences is 
forced to be always strictly linear.
Thus, $\nu$ must not be set to a value too small
to enable reasonably flexible time alignments.
To achieve orthogonality while enabling $\nu$ to be a moderately greater value, we propose 
introducing another loss to constrain ${\bf A}$, which we call
the orthogonal attention loss (OAL):
\begin{align}
\mathcal{L}_{\mathsf{oal}} = 
{\textstyle \frac{1}{\Nsrc^2}}
\|\Vec{W}_{\Nsrc\times \Nsrc}(\rho)\odot (\Vec{A}\Vec{A}^{\mathsf T})\|_1.
\end{align}

\subsection{Training loss}
\label{subsec:training}

Given examples of parallel utterances, 
the total training loss for the ConvS2S-VC model to be minimized is given as
\begin{align}
\mathcal{L} = 
\mathbb{E}_{\Xsrc,\Xtrg}
\left\{
\mathcal{L}_{\mathsf{dec}}
+
\lambda_{\mathsf{r}}
\mathcal{L}_{\mathsf{rec}}
+
\lambda_{\mathsf{d}}
\mathcal{L}_{\mathsf{dal}}
+
\lambda_{\mathsf{o}}
\mathcal{L}_{\mathsf{oal}}
\right\},
\end{align}
where 
$\mathbb{E}_{\Xsrc,\Xtrg}\{\cdot\}$ is 
the sample mean over all the training examples and 
$\lambda_{\mathsf{r}}\ge 0$, $\lambda_{\mathsf{d}}\ge 0$ and $\lambda_{\mathsf{o}}\ge 0$
are regularization parameters, which
weigh the importances of $\mathcal{L}_{\mathsf{rec}}$, 
$\mathcal{L}_{\mathsf{dal}}$ and $\mathcal{L}_{\mathsf{oal}}$
relative to $\mathcal{L}_{\mathsf{dec}}$.

\vspace{-2ex}
\subsection{Conversion process}
\label{subsec:conversion}
At test time, we can convert a source speech feature sequence $\Vec{X}$ via the following recursion:

\begin{algorithmic}
\STATE $[\Vec{K};\Vec{V}] = \mathsf{SrcEnc}(\Vec{X})$, ${\Vec{Y}} \leftarrow \Vec{0}$
\FOR{$m=1$ to $M'$}
\STATE $\Vec{Q} = \mathsf{TrgEnc}({\Vec{Y}})$
\STATE $\Vec{A} = \mathsf{softmax}\big({\Vec{K}^{\mathsf T}\Vec{Q}}/{\sqrt{D'}}\big)$
\STATE $\Vec{R} = \Vec{VA}$
\STATE $\Vec{Y} = \mathsf{TrgDec}(\Vec{R})$
\STATE $\Vec{Y} \leftarrow [\Vec{0},\Vec{Y}]$
\ENDFOR
\STATE \textbf{return} $\Vec{Y}$
\end{algorithmic}

\begin{figure}[t!]
\centering
\begin{minipage}{.49\linewidth}
\centerline{\includegraphics[width=.99\linewidth]{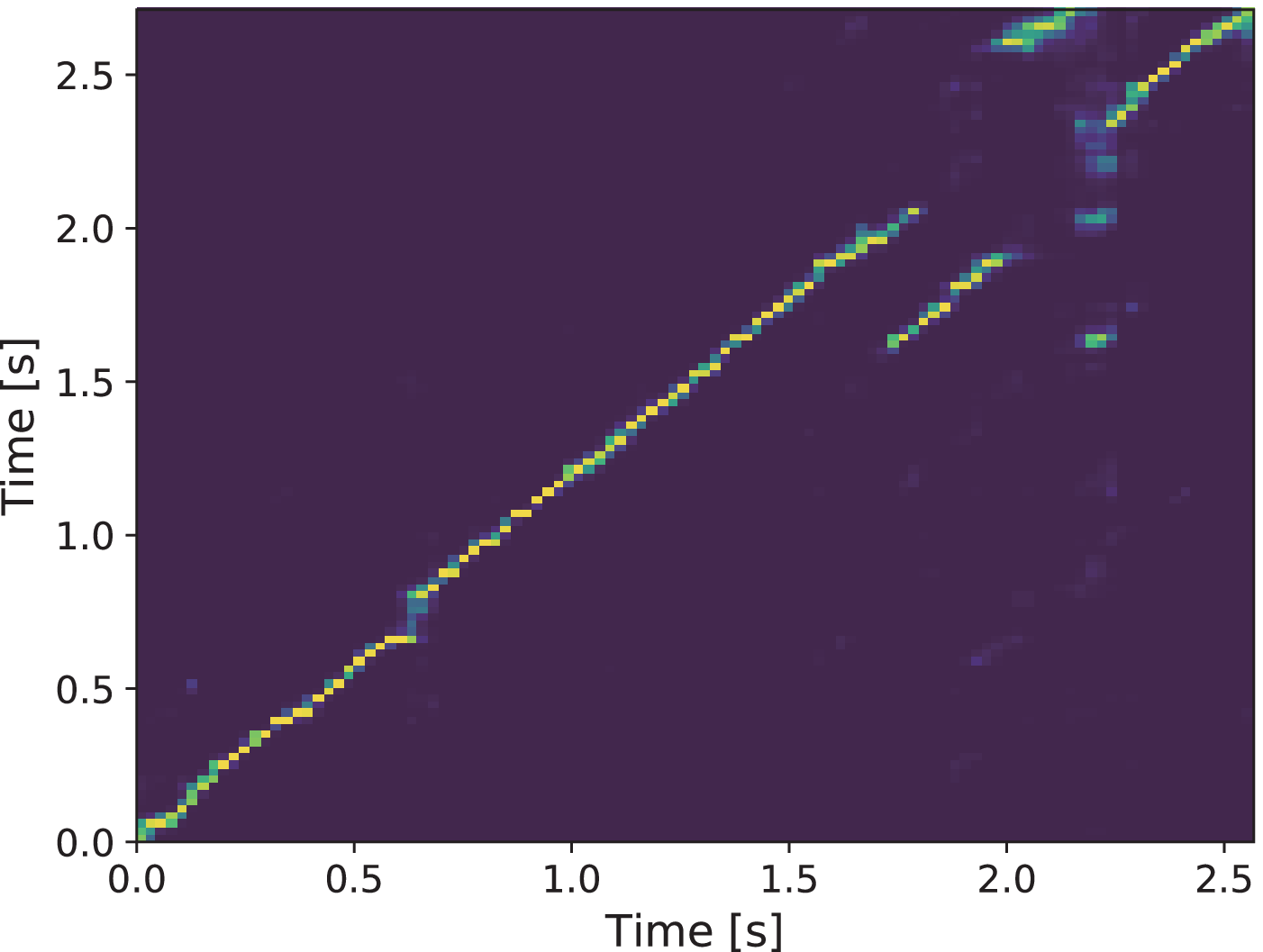}}
\end{minipage}
\begin{minipage}{.49\linewidth}
\centerline{\includegraphics[width=.99\linewidth]{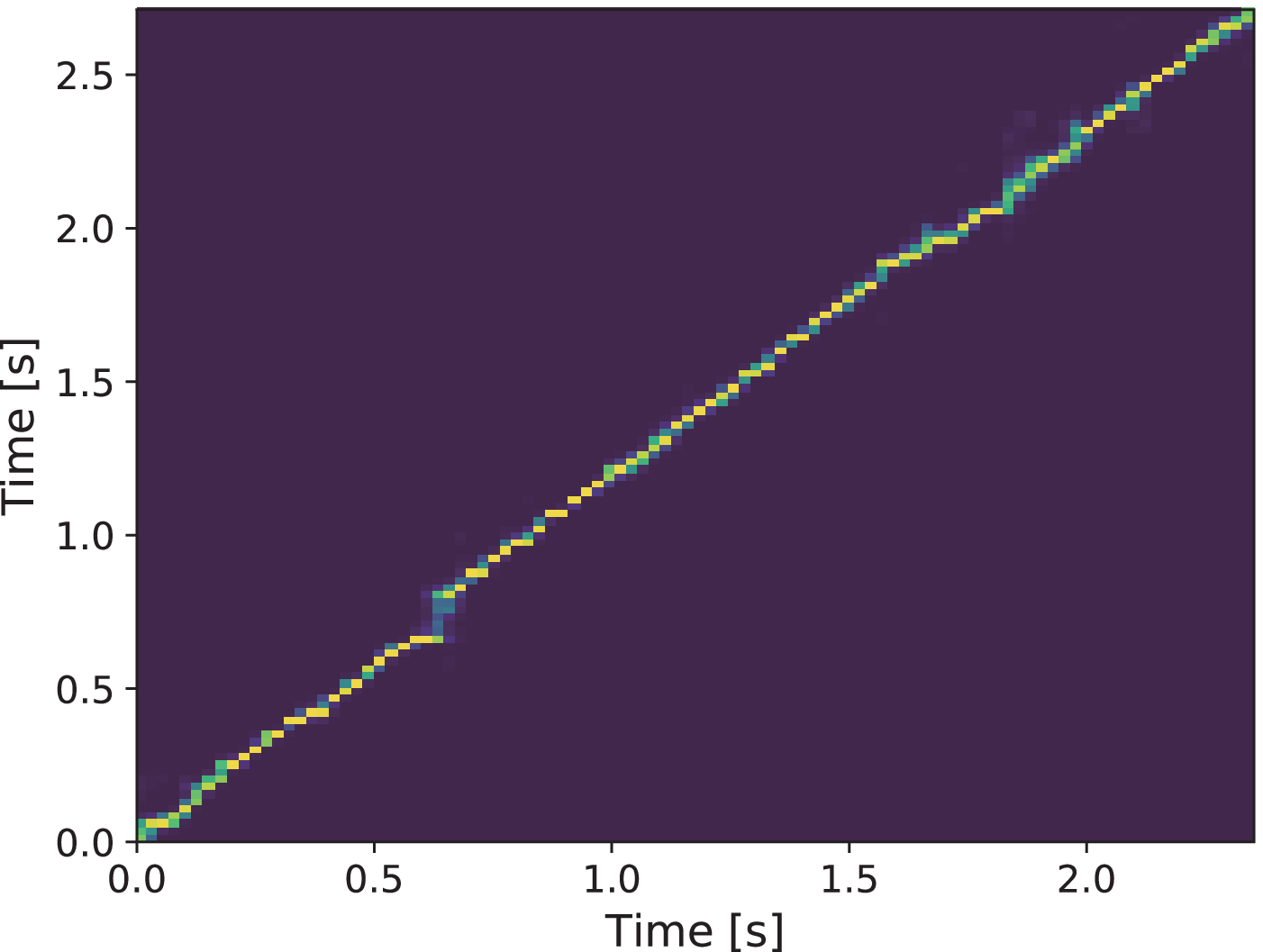}}
\end{minipage}
\vspace{-1ex}
\caption{
Attention matrices predicted  
without (left) and with (right) forward attention.}
\label{fig:forward}
\vspace{-1ex}
\end{figure}

However,
as \reffig{forward} shows,
it transpired that with this algorithm
the attended time point 
does not always move forward monotonically and continuously 
at test time  
and can occasionally 
become stuck at the same time point
or suddenly jump to a distant time point
even though the diagonal and orthogonal losses
are considered in training.
To assist the attended point to move forward monotonically and continuously, 
we limit the paths through which the attended point is allowed to move by forcing 
the attentions to the time points
distant from the peak of the attention distribution obtained at the previous time step
to zeros. This can be implemented for instance as follows:

\begin{algorithmic}
\STATE $[\Vec{K};\Vec{V}] = \mathsf{SrcEnc}(\Vec{X})$, ${\Vec{Y}} \leftarrow \Vec{0}$
\FOR{$m=1$ to $M$}
\STATE $\Vec{Q} = \mathsf{TrgEnc}({\Vec{Y}})$
\STATE $\Vec{A} = \mathsf{softmax}\big({\Vec{K}^{\mathsf T}\Vec{Q}}/{\sqrt{D'}}\big)$
\IF{$m>1$}
\STATE $\Vec{a} = \Vec{A}_{1:N,m}$
\STATE $\Vec{a}_{1:{\rm max}(1,\hat{n}-N_0)} = 0$, $\Vec{a}_{{\rm min}(\hat{n}+N_1,N):N} = 0$
\STATE $\Vec{a}\leftarrow \Vec{a}/{\rm sum}(\Vec{a})$
\STATE $\Vec{A} \leftarrow [\Vec{A}_{1:N,1:m-1},\Vec{a}]$
\ENDIF
\STATE $\hat{n} = \Argmax_{n}\Vec{A}_{n,m}$
\STATE $\Vec{R} = \Vec{VA}$
\STATE $\Vec{Y} = \mathsf{TrgDec}(\Vec{R})$
\STATE $\Vec{Y} \leftarrow [\Vec{0},\Vec{Y}]$
\ENDFOR
\STATE \textbf{return} $\Vec{Y}$
\end{algorithmic}

\noindent
where ${\rm sum}(\cdot)$ denotes the sum of all the elements in a vector.
Note that we set $N_0$ and $N_1$ at the nearest integers that 
correspond to 
$160$[ms] and $320$[ms], respectively.
\reffig{forward} shows an example of how attention matrices look different 
when this procedure has been undertaken.
After we obtain $\Vec{R}$ with the above algorithm,
we can use the target reconstructor to compute 
$\widetilde{\Vec{Y}} = \mathsf{TrgRec}(\Vec{R})$
and use it instead of $\Vec{Y}$ 
as the feature sequence of the converted speech.

Once $\Vec{Y}$ or $\widetilde{\Vec{Y}}$ has been obtained, 
we adjust
the mean and variance of the generated feature sequence 
so that they match the pretrained 
mean and variance of the feature vectors of the target speaker.
We can then generate a time-domain signal using the WORLD vocoder or any 
recently developed neural vocoder \cite{vandenOord2016short,Tamamori2017short,Kalchbrenner2018short,Mehri2016short,Jin2018short, vandenOord2017short, Ping2019short,Prenger2018short,Kim2018short,Wang2018short,Tanaka2018short}.
Note that in the following experiments, 
we chose to use $\widetilde{\Vec{Y}}$
for final waveform generation
as it resulted in better-sounding speech.

\subsection{Real-Time System Design}
\label{subsec:real-time}

Real-time requirements must be considered when building VC systems. 
If we want our model to work in real-time, 
first, we must not allow 
the source encoder to use future information as with the target encoder and decoder
during training. 
This requirement can easily be implemented by 
constraining the convolution layers in the source encoder (and the target reconstructor, 
if we assume it is used to generate the converted feature sequence) to be causal. 
Another point we must consider is that
the speaking rate and rhythm of input speech cannot be changed drastically at test time. 
One simple way of keeping them unchanged is to set 
$\Vec{A}$ to an identity matrix. 
In this way, the autoregressive recursion will be no longer needed 
and the conversion can be performed in a sliding-window fashion as: 

\begin{algorithmic}
\STATE $[\Vec{K};\Vec{V}] = \mathsf{SrcEnc}(\Vec{X})$
\STATE $\Vec{Y} = \mathsf{TrgDec}(\Vec{V})$~or~$\Vec{Y} = \mathsf{TrgRec}(\Vec{V})$
\STATE \textbf{return} $\Vec{Y}$
\end{algorithmic}

\noindent
We will show later how these modifications can affect the VC performance.
Note that 
even under this setting, 
the ability to learn and apply conversion rules 
that capture long-term dependencies 
is still effective.

\subsection{Impact of Batch Normalization}

As mentioned earlier, using fully convolutional architectures allows the use of 
batch normalization for all the hidden layers in the networks, which 
is not straightforward for architectures including recurrent modules.
One benefit of using batch normalization layers is that it enables the networks to use a higher learning rate without vanishing or exploding gradients. It is also believed to help regularize the networks such that it is easier to generalize and mitigate overfitting. 
The effect of batch normalization will be verified experimentally in \refsec{experiments}.

\section{Many-to-Many ConvS2S-VC}
\label{sec:ConvS2S-VC2}

\subsection{Model and Training Loss}
\label{subsec:multidomain}

We now describe an extension of the ConvS2S model
that enables many-to-many VC.
Here, the idea is to use a single model to achieve mappings among multiple speakers.
The model consists of the same set of the networks as the pairwise model.
The only difference is that 
each network takes a speaker index as an additional input.

Let $\Vec{X}^{(1)},\ldots,\Vec{X}^{(K)}$ 
be examples of the acoustic feature sequences of speech in different speakers reading the same sentence.
Given a single pair of parallel utterances $\Vec{X}^{(k)}$ and $\Vec{X}^{(k')}$, 
where $k$ and $k'$ denote the source and target speaker indices 
(integers),
the source encoder takes $\Vec{X}^{(k)}$ and the source speaker index $k$
as the inputs and produces two internal vector sequences $\Vec{K}^{(k)}, \Vec{V}^{(k)}$,
whereas the target encoder 
takes $\Vec{X}^{(k')}$ and the target speaker index $k'$
as the inputs and produces an internal vector sequence $\Vec{Q}^{(k')}$:
\begin{align}
[\Vec{K}^{(k)};\Vec{V}^{(k)}] &= \mathsf{SrcEnc}(\Vec{X}^{(k)}, k),
\label{eq:multidomain_srcenc}
\\
\Vec{Q}^{(k')} &= \mathsf{TrgEnc}(\Vec{X}^{(k')},k').
\end{align}
The attention matrix $\Vec{A}^{(k,k')}$ 
and the time-warped version of $\Vec{V}^{(k)}$ are
then computed using 
$\Vec{K}^{(k)}$ and $\Vec{Q}^{(k')}$:
\begin{align}
\Vec{A}^{(k,k')} &= \mathsf{softmax}_{n}
\big(
\Vec{K}^{(k)}{}^{\mathsf T} \Vec{Q}^{(k')}\big/\sqrt{D'}
\big),\\
\Vec{R}^{(k,k')} &= \Vec{V}^{(k)} \Vec{A}^{(k,k')}.
\end{align}
The outputs of the reconstructor 
and decoder
given the input $\Vec{R}^{(k,k')}$
with target speaker conditioning
are finally given as
\begin{align}
\widetilde{\Vec{Y}}{}^{(k,k')} &= 
\mathsf{TrgRec}(\Vec{R}^{(k,k')},k'),\\
\Vec{Y}^{(k,k')} &= \mathsf{TrgDec}(\Vec{R}^{(k,k')},k').
\end{align}
The loss functions to be minimized
given this single training example are given as
\begin{align}
\mathcal{L}_{\mathsf{dec}}^{(k,k')} &=
{\textstyle 
\frac{1}{N_{k'}}
}
\| \Vec{Y}_{:,1:N_{k'}-1}^{(k,k')} - \Vec{X}_{:,2:N_{k'}}^{(k')} \|_1,
\label{eq:decloss_multidomain}
\\
\mathcal{L}_{\mathsf{dal}}^{(k,k')} &= 
{\textstyle \frac{1}{N_{k}N_{k'}}}
\|\Vec{W}_{N_{k}\times N_{k'}}(\nu)\odot \Vec{A}^{(k,k')}\|_1,
\\
\mathcal{L}_{\mathsf{oal}}^{(k,k')} &= 
{\textstyle \frac{1}{N_{k}{}^2}}
\|\Vec{W}_{N_{k}\times N_{k}}(\rho)\odot 
(\Vec{A}^{(k,k')}
\Vec{A}^{(k,k')}{}^{\mathsf T})
\|_1,
\\
\mathcal{L}_{\mathsf{rec}}^{(k,k')} &= 
{\textstyle \frac{1}{N_{k'}}}
\|
\widetilde{\Vec{Y}}{}^{(k,k')}
-
{\Vec{X}}^{(k')}
\|_1.
\label{eq:recloss_multidomain}
\end{align}
With the above model,
the case where $k=k'$ would be reasonable to also consider.
Minimizing this loss corresponds to
ensuring that
the input feature sequence $\Vec{X}^{(k)}$ 
will remain unchanged when the source and target speakers are the same.
We call this loss the ``identity mapping loss (IML)''.
The effect given by this loss will be shown later.
Hence, 
the total training loss to be minimized becomes
\begin{align}
&\mathcal{L} =
\sum_{k,k'\neq k}
\mathbb{E}_{\Vec{X}^{(k)}\!,\Vec{X}^{(k')}}
\!
\big\{
\mathcal{L}_{\mathsf{all}}^{(k,k')}
\big\}
+
\lambda_{\mathsf{i}}
\sum_{k}
\mathbb{E}_{\Vec{X}^{(k)}}
\!
\big\{
\mathcal{L}_{\mathsf{all}}^{(k,k)}
\big\},
\nonumber\\
&\mathcal{L}_{\mathsf{all}}^{(k,k')}=
\mathcal{L}_{\mathsf{dec}}^{(k,k')}
+
\lambda_{\mathsf{r}}
\mathcal{L}_{\mathsf{rec}}^{(k,k')}
+
\lambda_{\mathsf{d}}
\mathcal{L}_{\mathsf{dal}}^{(k,k')}
+
\lambda_{\mathsf{o}}
\mathcal{L}_{\mathsf{oal}}^{(k,k')},
\end{align}
where
$\mathbb{E}_{\Vec{X}^{(k)}\!,\Vec{X}^{(k')}}[\cdot]$
and $\mathbb{E}_{\Vec{X}^{(k)}}[\cdot]$
denote the sample means over all the training examples of parallel utterances in speakers $k$ and $k'$,
and $\lambda_{\mathsf{i}}\ge 0$ is a regularization parameter, which weighs the importance of 
the IML. 

\subsection{Conditional Batch Normalization}
\label{subsec:dbn}

\begin{figure}[t!]
\centering
\begin{minipage}{.49\linewidth}
  \centerline{\includegraphics[width=.99\linewidth]{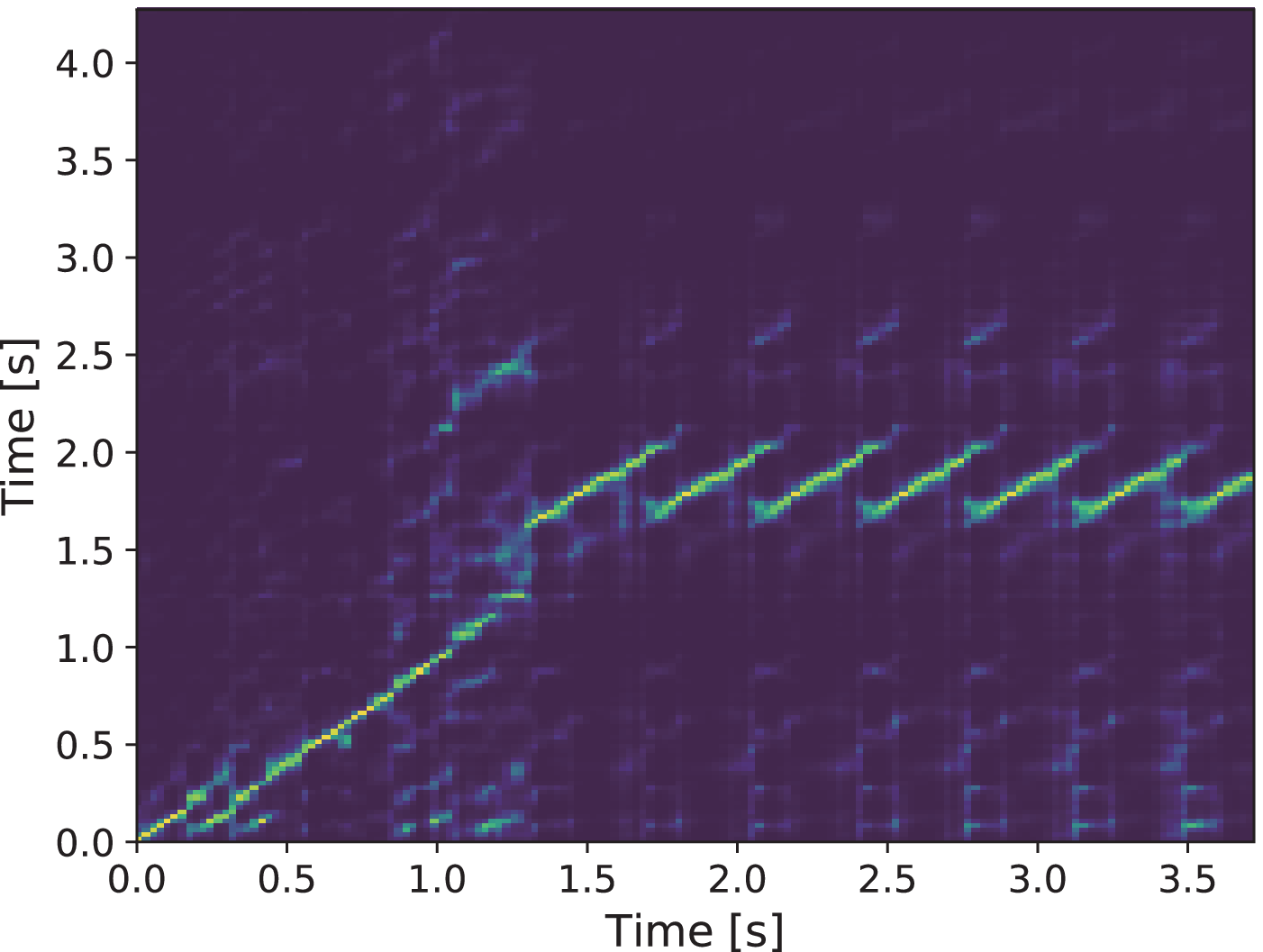}}
\end{minipage}
\begin{minipage}{.49\linewidth}
  \centerline{\includegraphics[width=.99\linewidth]{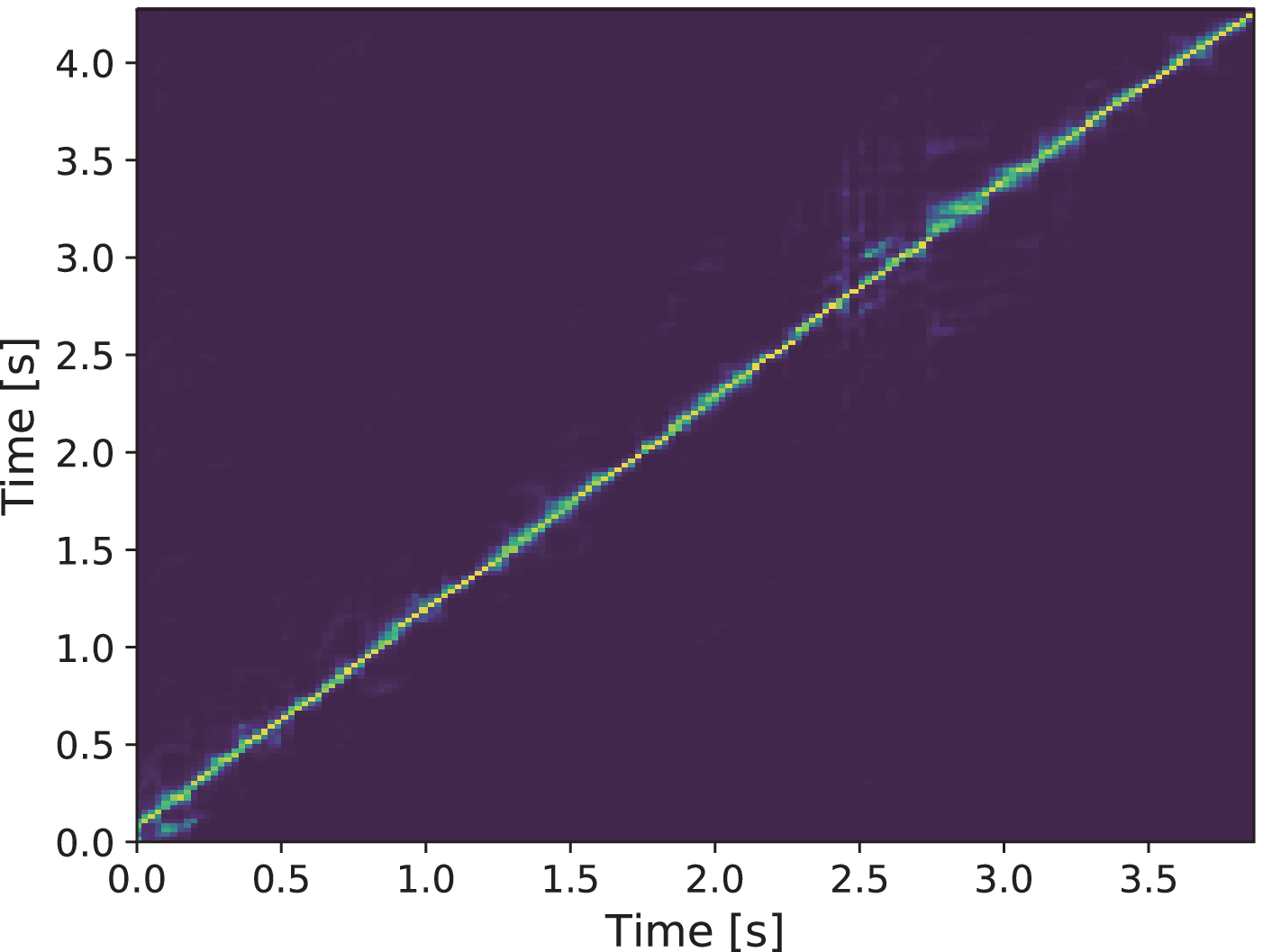}}
\end{minipage}
\vspace{-1ex}
\caption{Examples of the attention matrices predicted from 
test input female speech using the many-to-many model: 
with batch normalization (left) and with conditional batch normalization (right).}
\label{fig:multidomain}
\vspace{-1ex}
\end{figure}

The left figure in 
\reffig{multidomain} shows 
the attention matrix predicted from input female speech
using the many-to-many model with regular
batch normalization layers.
As this example shows, attention matrices predicted by the many-to-many model
tended to become blurry, mostly resulting in unintelligible speech.
We conjecture that this was caused by the fact that 
the distributions of the inputs to the hidden layers can 
change in accordance with the source and/or target speakers.
To normalize layer input distributions on a speaker-dependent basis,
we propose 
using conditional batch normalization layers 
for the many-to-many model. 
Each element $y_{b,d,n}$ of the output of a
regular batch normalization layer 
$\Vec{Y} = \mathsf{B}(\Vec{X})$
is defined as
$y_{b,d,n} = 
\gamma_{d}
\frac{x_{b,d,n}-\mu_{d}(\Vec{X})}{\sigma_{d}(\Vec{X})}
+\beta_d$,
where $\Vec{X}$ denotes the layer input given by a three-way array with batch, channel, and time axes, 
$x_{b,d,n}$ denotes its $(b,d,n)$-th element, 
$\mu_d(\Vec{\Vec{X}})$ and $\sigma_d(\Vec{X})$ denote the mean and standard deviation of 
the $d$-th channel components of $\Vec{X}$ 
computed along the batch and time axes, and $\Vec{\gamma}=[\gamma_1,\ldots,\gamma_D]$ 
and $\Vec{\beta}=[\beta_1,\ldots,\beta_D]$ denote the parameters to be learned.
In contrast, 
the output of a
conditional batch normalization layer 
$\Vec{Y} = \mathsf{B}^k(\Vec{X})$
is defined as 
$y_{b,d,n} = 
\gamma_{d}^{k}
\frac{x_{b,d,n}-\mu_{d}(\Vec{X})}{\sigma_{d}(\Vec{X})}
+\beta_d^{k}$, where the only difference is that 
the parameters $\Vec{\gamma}^k=[\gamma_1^k,\ldots,\gamma_D^k]$ 
and $\Vec{\beta}^k=[\beta_1^k,\ldots,\beta_D^k]$ are conditioned on speaker $k$.
Note that a similar idea, 
called the conditional instance normalization, has been introduced to modify 
the instance normalization process for image style transfer 
\cite{Dumoulin2017short} and non-parallel VC \cite{Kaneko2019short_starganvc2}.

\subsection{Any-to-many Conversion}
\label{subsec:many2one}

With the models presented above,
the source speaker must be known and specified during both training and inference.
However, there can be certain situations where the source speaker 
is unknown or arbitrary.
We call VC tasks in such scenarios any-to-one or any-to-many VC.
Our many-to-many model can be modified to
handle any-to-many VC tasks
by not allowing the source encoder to take 
the source speaker index $k$ as an input 
at both training and test time.
The modified version 
can be formulated by simply
replacing \refeq{multidomain_srcenc} in the many-to-many model with
\begin{align}
[\Vec{K}^{(k)};\Vec{V}^{(k)}] &= \mathsf{SrcEnc}(\Vec{X}^{(k)}).
\label{eq:many2one_srcenc}
\end{align}

\begin{table*}[t!]
\centering
\caption{Notations for network architecture descriptions}
\begin{tabular}{l|l}
\thline
Notation&Meaning\\\hline
$\Vec{X}$ (bold symbol)&two-way array with channel and time axes\\
$(f \circ g)(\Vec{X})$&function composition $f(g(\Vec{X}))$\\
$(\bigcirc_{n=0}^{N} f_n)(\Vec{X})$&multiple compositions $(f_N \circ \cdots \circ f_1 \circ f_0)(\Vec{X})$\\
$\mathsf{RL}_{\kappa\star \delta}^{o\leftarrow i}(\Vec{X})$&1D regular (non-causal) convolution ($\kappa$: kernel size, $\delta$: dilation factor, $i$: input channel size, $o$: output channel size)\\
$\mathsf{CL}_{\kappa\star \delta}^{o\leftarrow i}(\Vec{X})$&1D causal convolution ($\kappa$: kernel size, $\delta$: dilation factor, $i$: input channel size, $o$: output channel size)\\
$\mathsf{B}(\Vec{X})$&normalization (BN or IN)\\
$\mathsf{B}^{k}(\Vec{X})$&conditional normalization (CBN or CIN) conditioned on speaker $k$\\
$\mathsf{ResRGLU}_{\kappa\star \delta}^{o\leftarrow i}(\Vec{X})$&
$\mathsf{B}(\mathsf{RL}_{\kappa\star \delta}^{o\leftarrow i}(\Vec{X}))\odot \mathsf{sigmoid}(\mathsf{B}(\mathsf{RL}_{\kappa\star \delta}^{o\leftarrow i}(\Vec{X}))) + \Vec{X}_{1:o,:}$\\
$\mathsf{ResCGLU}_{\kappa\star \delta}^{o\leftarrow i}(\Vec{X})$&
$\mathsf{B}(\mathsf{CL}_{\kappa\star \delta}^{o\leftarrow i}(\Vec{X}))\odot \mathsf{sigmoid}(\mathsf{B}(\mathsf{CL}_{\kappa\star \delta}^{o\leftarrow i}(\Vec{X}))) + \Vec{X}_{1:o,:}$\\
$\mathsf{ResRGLU}^{k}{}_{\kappa\star \delta}^{o\leftarrow i}(\Vec{X})$&
$\mathsf{B}^{k}(\mathsf{RL}_{\kappa\star \delta}^{o\leftarrow i}(\Vec{X}))\odot \mathsf{sigmoid}(\mathsf{B}^{k}(\mathsf{RL}_{\kappa\star \delta}^{o\leftarrow i}(\Vec{X}))) + \Vec{X}_{1:o,:}$\\
$\mathsf{ResCGLU}^{k}{}_{\kappa\star \delta}^{o\leftarrow i}(\Vec{X})$&
$\mathsf{B}^{k}(\mathsf{CL}_{\kappa\star \delta}^{o\leftarrow i}(\Vec{X}))\odot \mathsf{sigmoid}(\mathsf{B}^{k}(\mathsf{CL}_{\kappa\star \delta}^{o\leftarrow i}(\Vec{X}))) + \Vec{X}_{1:o,:}$\\
$\mathsf{drop}(\Vec{X})$&dropout with ratio 0.1\\ 
$\mathsf{embed}^{o}(k)$&retrieving an $o$-dimensional embedding vector
from an integer $k$
\\
$\mathsf{A}_{\Vec{Z}}(\Vec{X})$&appending a broadcast version of $\Vec{Z}$ (expanded along the time axis) to $\Vec{X}$ along the channel dimension\\
\thline
\end{tabular}
\label{tab:notations}
\end{table*}

\begin{table*}[t!]
\centering
\caption{Network architectures of pairwise and many-to-many models}
\begin{tabular}{l|l|l}
\thline
Model&Network&Architecture\\\hline
\multirow{4}{*}{
\parbox{1.7cm}{pairwise}
}
&
$\mathsf{SrcEnc}(\Vec{X})$&$
(\mathsf{RL}_{1\star 1}^{512 \leftarrow 256} \circ
(\bigcirc_{l'=0}^{2} (\bigcirc_{l=0}^{3} \mathsf{ResRGLU}_{5\star 3^l}^{256 \leftarrow 256}))
\circ \mathsf{B} \circ \mathsf{RL}_{1\star 1}^{256 \leftarrow 93}\circ \mathsf{drop})(\Vec{X})$\\
&
$\mathsf{TrgEnc}(\Vec{Y})$&$
(\mathsf{CL}_{1\star 1}^{256 \leftarrow 256} \circ
(\bigcirc_{l'=0}^{2} (\bigcirc_{l=0}^{3} \mathsf{ResCGLU}_{3\star 3^l}^{256 \leftarrow 256}))
\circ \mathsf{B} \circ \mathsf{CL}_{1\star 1}^{256 \leftarrow 93}\circ \mathsf{drop})(\Vec{Y})$\\
&
$\mathsf{TrgRec}(\Vec{R})$&$
(\mathsf{RL}_{1\star 1}^{93 \leftarrow 256} \circ
(\bigcirc_{l'=0}^{2} (\bigcirc_{l=0}^{3} \mathsf{ResRGLU}_{5\star 3^l}^{256 \leftarrow 256}))
\circ \mathsf{B} \circ \mathsf{RL}_{1\star 1}^{256 \leftarrow 256}\circ \mathsf{drop})(\Vec{R})$\\
&
$\mathsf{TrgDec}(\Vec{R})$&$
(\mathsf{CL}_{1\star 1}^{93 \leftarrow 256} \circ
(\bigcirc_{l'=0}^{2} (\bigcirc_{l=0}^{3} \mathsf{ResCGLU}_{3\star 3^l}^{256 \leftarrow 256}))
\circ \mathsf{B} \circ \mathsf{CL}_{1\star 1}^{256 \leftarrow 256}\circ \mathsf{drop})(\Vec{R})$\\
\hline
\multirow{6}{*}{
\parbox{1.7cm}{many-to-many\\(batch)}
}
&
$\mathsf{SrcEnc}(\Vec{X},k)$&$
(\mathsf{RL}_{1\star 1}^{1024 \leftarrow 544} \circ \mathsf{A}_{\Vec{Z}} \circ
(\bigcirc_{l'=0}^{2} (\bigcirc_{l=0}^{3} (\mathsf{ResRGLU}^{k}{}_{5\star 3^l}^{512 \leftarrow 544} \circ \mathsf{A}_{\Vec{Z}})))
\circ \mathsf{B}^{k} \circ \mathsf{RL}_{1\star 1}^{512 \leftarrow 125}\circ \mathsf{A}_{\Vec{Z}} \circ \mathsf{drop})(\Vec{X})$\\
&
$\mathsf{TrgEnc}(\Vec{Y},{k'})$&$
(\mathsf{CL}_{1\star 1}^{512 \leftarrow 544} \circ \mathsf{A}_{\Vec{Z}'} \circ
(\bigcirc_{l'=0}^{2} (\bigcirc_{l=0}^{3} (\mathsf{ResCGLU}^{k'}{}_{3\star 3^l}^{512 \leftarrow 544} \circ \mathsf{A}_{\Vec{Z}'} )))
\circ \mathsf{B}^{k'} \circ \mathsf{CL}_{1\star 1}^{512 \leftarrow 125}\circ \mathsf{A}_{\Vec{Z}'} \circ \mathsf{drop})(\Vec{Y})$\\
&
$\mathsf{TrgRec}(\Vec{R},{k'})$&$
(\mathsf{RL}_{1\star 1}^{93 \leftarrow 544} \circ \mathsf{A}_{\Vec{Z}'} \circ
(\bigcirc_{l'=0}^{2} (\bigcirc_{l=0}^{3} (\mathsf{ResRGLU}^{k'}{}_{5\star 3^l}^{512 \leftarrow 544} \circ \mathsf{A}_{\Vec{Z}'})))
\circ \mathsf{B}^{k'} \circ \mathsf{RL}_{1\star 1}^{512 \leftarrow 544}\circ \mathsf{A}_{\Vec{Z}'} \circ \mathsf{drop})(\Vec{R})$\\
&
$\mathsf{TrgDec}(\Vec{R},{k'})$&$
(\mathsf{CL}_{1\star 1}^{93 \leftarrow 544} \circ \mathsf{A}_{\Vec{Z}'} \circ
(\bigcirc_{l'=0}^{2} (\bigcirc_{l=0}^{3} (\mathsf{ResCGLU}^{k'}{}_{3\star 3^l}^{512 \leftarrow 544} \circ \mathsf{A}_{\Vec{Z}'} )))
\circ \mathsf{B}^{k'} \circ \mathsf{CL}_{1\star 1}^{512 \leftarrow 544}\circ \mathsf{A}_{\Vec{Z}'} \circ \mathsf{drop})(\Vec{R})$ \\
&&where $\Vec{Z}=\mathsf{embed}^{32}(k)$ and $\Vec{Z}'=\mathsf{embed}^{32}(k')$\\
\hline
\multirow{2}{*}{
any-to-many
}
&
$\mathsf{SrcEnc}(\Vec{X})$&$
(\mathsf{RL}_{1\star 1}^{1024 \leftarrow 512} \circ 
(\bigcirc_{l'=0}^{2} (\bigcirc_{l=0}^{3} \mathsf{ResRGLU}_{5\star 3^l}^{512 \leftarrow 512} ))
\circ \mathsf{B} \circ \mathsf{RL}_{1\star 1}^{512 \leftarrow 93}\circ \mathsf{drop})(\Vec{X})$\\
&others&same as above\\
\hline
\multirow{3}{*}{
\parbox{1.7cm}{many-to-many\\(real-time)}
}
&
$\mathsf{SrcEnc}(\Vec{X},k)$&$
(\mathsf{CL}_{1\star 1}^{1024 \leftarrow 544} \circ \mathsf{A}_{\Vec{Z}} \circ
(\bigcirc_{l'=0}^{2} (\bigcirc_{l=0}^{3} (\mathsf{ResCGLU}^{k}{}_{3\star 3^l}^{512 \leftarrow 544} \circ \mathsf{A}_{\Vec{Z}})))
\circ \mathsf{B}^{k} \circ \mathsf{CL}_{1\star 1}^{512 \leftarrow 125}\circ \mathsf{A}_{\Vec{Z}} \circ \mathsf{drop})(\Vec{X})$\\
&
$\mathsf{TrgRec}(\Vec{R},{k'})$&$
(\mathsf{CL}_{1\star 1}^{93 \leftarrow 544} \circ \mathsf{A}_{\Vec{Z}'} \circ
(\bigcirc_{l'=0}^{2} (\bigcirc_{l=0}^{3} (\mathsf{ResCGLU}^{k'}{}_{3\star 3^l}^{512 \leftarrow 544} \circ \mathsf{A}_{\Vec{Z}'})))
\circ \mathsf{B}^{k'} \circ \mathsf{CL}_{1\star 1}^{512 \leftarrow 544}\circ \mathsf{A}_{\Vec{Z}'} \circ \mathsf{drop})(\Vec{R})$\\
&
$\mathsf{TrgDec}(\Vec{R},{k'})$
&same as above\\
\thline
\end{tabular}
\label{tab:architectures}
\end{table*}

\section{Experiments}
\label{sec:experiments}

\subsection{Experimental Settings}
\label{subsec:expcond}

To evaluate 
the effects of the ideas presented in 
Sections \ref{sec:ConvS2S-VC} and \ref{sec:ConvS2S-VC2}, 
we conducted 
objective and subjective evaluation experiments involving
a speaker identity conversion task. 
For the experiment, we used the CMU Arctic database \cite{Kominek2004short},
which consists of recordings of 1132 phonetically balanced English utterances  
spoken by four US English speakers. 
We used all the speakers 
(clb (female), 
bdl (male),
slt (female), and 
rms (male)) for training and evaluation.
Thus, in total 
there were 12 different combinations of source and target speakers.
The audio files for each speaker were manually
divided into 1000 and 132 files, which were provided
as training and evaluation sets, respectively. 
All the speech signals were sampled at 16 kHz. 
As already detailed in \refsubsec{feature},
for each utterance, 
the spectral envelope,
log $F_0$, 
coded aperiodicity,
and voiced/unvoiced information
were extracted every 8 ms 
using the
WORLD analyzer \cite{Morise2016short}. 
Then,
28 MCCs were extracted from 
each spectral envelope using the Speech Processing Toolkit (SPTK) \cite{pysptk_url}.
The reduction factor $r$ was set to $3$.
Hence, the dimension of the acoustic feature was $D=(28+3)\times 3=93$.

\subsection{Network architectures}
\label{subsec:netarch}

We use the notations in Tab. \ref{tab:notations}
to describe the network architectures. 
The architectures of all the networks in the pairwise and 
many-to-many models are 
detailed in Tab. \ref{tab:architectures}. 
Note that in Tab. \ref{tab:architectures} the layer index is omitted for simplicity of notation and
each layer has a different set of free parameters even though the same symbol is used. 

\subsection{Hyperparameter Settings}

$\lambda_{\mathsf{r}}$, $\lambda_{\mathsf{d}}$, $\lambda_{\mathsf{o}}$, and
$\lambda_{\mathsf{i}}$
were set at 1, 2000, 2000, and 1, respectively.
$\nu$ and $\rho$ were set at 
0.3 and 0.3 for both the pairwise and many-to-many models.
The $L_1$ norm $\|\Vec{X}\|_1$ used in 
(\ref{eq:decloss_pairwise}), 
(\ref{eq:recloss_pairwise}),
(\ref{eq:decloss_multidomain}), and  
(\ref{eq:recloss_multidomain})
was defined as a weighted norm
\begin{align}
\|\Vec{X}\|_1 =
\sum_{n=1}^{N}
\frac{1}{r}
\sum_{j=1}^{r}
\sum_{i=1}^{31} 
\alpha_i
|x_{ij,n}|,
\nonumber
\end{align}
where 
$x_{1j,n},\ldots,x_{28j,n}$,
$x_{29j,n}$,
$x_{30j,n}$ and $x_{31j,n}$
denote the entries 
of $\Vec{X}$ corresponding to 
the 28 MCCs, log $F_0$, coded aperiodicity
and voiced/unvoiced indicator 
at time $n$,
and the weights 
were set at
$\alpha_1=\cdots=\alpha_{28}=\frac{1}{28}$,
$\alpha_{29}=\frac{1}{10}$, and
$\alpha_{30}=\alpha_{31}=\frac{1}{50}$, respectively. 

All the networks were trained simultaneously with random initialization.
Adam optimization \cite{Kingma2015short} was used for model training where 
the mini-batch size was 16 and 25,000 iterations were run. 
The learning rate and the exponential decay rate
for the first moment 
for Adam were set at 0.00015 and 0.9.

\subsection{Objective Performance Measures}

The test dataset consists 
of speech samples of each speaker reading the same sentences.
Thus, 
the quality of a converted feature sequence can be assessed by
comparing it with the feature sequence of the reference utterance.

\subsubsection{Mel-cepstral distortion (MCD)}
Given two mel-cepstra,
$\hat{\Vec{x}} = [\hat{x}_1,\ldots,\hat{x}_{28}]^{\mathsf T}$ 
and $\Vec{x} = [x_1,\ldots,x_{28}]^{\mathsf T}$,
we can use the mel-cepstral distortion (MCD):
\begin{align}
{\rm MCD [dB]}=
\frac{10}{\ln 10}
\sqrt{
2\sum_{i=2}^{28} (\hat{x}_i-x_i)^2
},
\end{align}
to measure their difference.
Here, we used the average of the MCDs 
taken along the 
DTW
path between converted and reference feature sequences
as the objective performance measure for each test utterance.

\subsubsection{Log $F_0$ Correlation Coefficient (LFC)}
To evaluate the $F_0$ contour of converted speech, 
we used the correlation coefficient between the predicted and target log $F_0$ contours \cite{Hermes1998short}
as the objective performance measure. 
Since the converted and reference utterances were not necessarily aligned in time, 
we computed the correlation coefficient after properly aligning them.
Here, we used 
the MCC sequences 
$\hat{\Vec{X}}_{1:28,1:N}$,
${\Vec{X}}_{1:28,1:M}$
of converted and reference utterances
to find phoneme-based alignment,
assuming that the predicted and reference MCCs 
at the corresponding frames were sufficiently close.
Given the log $F_0$ contours
$\hat{\Vec{X}}_{29,1:N}$, 
$\Vec{X}_{29,1:M}$
and 
the voiced/unvoiced indicator sequences
$\hat{\Vec{X}}_{31,1:N}$, 
$\Vec{X}_{31,1:M}$
of converted and reference utterances,
we first warp the time axis of 
$\hat{\Vec{X}}_{29,1:N}$ and $\hat{\Vec{X}}_{31,1:N}$
in accordance with the DTW path between the MCC sequences 
$\hat{\Vec{X}}_{1:28,1:N}$,
${\Vec{X}}_{1:28,1:M}$
of the two utterances 
and obtain their time-warped versions, 
$\tilde{\Vec{X}}_{29,1:M}$,
$\tilde{\Vec{X}}_{31,1:M}$.
We then extract the elements of 
$\tilde{\Vec{X}}_{29,1:M}$ and 
$\Vec{X}_{29,1:M}$ 
at all the time points corresponding to the voiced segments
such that 
$\{m|\tilde{\Vec{X}}_{31,m}\!=\!\Vec{X}_{31,m}\!=\!1\}$.
If we use 
$\tilde{\Vec{y}}=[\tilde{y}_1,\ldots,\tilde{y}_{M'}]$ and 
$\Vec{y}=[y_1,\ldots,y_{M'}]$ to
denote the vectors consisting of the elements extracted from 
$\tilde{\Vec{X}}_{29,1:M}$ and $\Vec{X}_{29,1:M}$, 
we can use the correlation coefficient between $\tilde{\Vec{y}}$ and $\Vec{y}$
\begin{align}
R = 
\frac{
\sum_{m'=1}^{M'} 
(\tilde{y}_{m'} - \tilde{\varphi}) 
({y}_{m'} - \varphi)
}{
\sqrt{
 \sum_{{m'}=1}^{M'}(\tilde{y}_{m'} - \tilde{\varphi})^2
}
\sqrt{
 \sum_{m'=1}^{M'}({y}_{m'} - {\varphi})^2
}
},
\end{align}
where $\tilde{\varphi} = \frac{1}{M'} \sum_{m'=1}^{M'} \tilde{y}_{m'}$ and 
$\varphi = \frac{1}{M'} \sum_{m'=1}^{M'} {y}_{m'}$,
to measure the similarity between the two log $F_0$ contours.
In the current experiment, 
we used the average of the correlation coefficients taken over all the test utterances 
as the objective performance measure for log $F_0$ prediction.
Thus, the closer it is to 1, the better the performance.
We call this measure the log $F_0$ correlation coefficient (LFC).

\subsubsection{Local Duration Ratio (LDR)}

To evaluate 
the speaking rate and the rhythm of converted speech, 
we used the local slopes of the DTW path between 
converted and reference utterances to determine the objective performance measure.
If the speaking rate and the rhythm of the two utterances are exactly the same,
all the local slopes should be 1. 
Hence, the better the conversion, the closer the local slopes become to 1.
To compute the local slopes, we undertook the following process.
Given the MCC sequences 
$\hat{\Vec{X}}_{1:28,1:N}$,
${\Vec{X}}_{1:28,1:M}$
of converted and reference utterances, 
we first performed DTW on 
$\hat{\Vec{X}}_{1:28,1:N}$ and ${\Vec{X}}_{1:28,1:M}$.
If we use $(p_1,q_1),\ldots,(p_j,q_j),\ldots,(p_J,q_J)$ 
to denote the obtained DTW path where $(p_1,q_1)=(1,1)$ and $(p_J,q_J)=(M,N)$,  
we computed the slope of the regression line fitted to 
the $33$ local consecutive points for each $j$:
\begin{align}
s_j = 
\frac{
\sum_{j'=j-16}^{j+16}
(p_{j'}-\bar{p}_j)(q_{j'}-\bar{q}_j)
}{
\sum_{j'=j-16}^{j+16}
(p_{j'}-\bar{p}_j)^2
},
\end{align}
where $\bar{p}_j=\frac{1}{33}\sum_{j'=j-16}^{j+16} p_{j'}$ and
$\bar{q}_j=\frac{1}{33}\sum_{j'=j-16}^{j+16} q_{j'}$, 
and then computed the median of $s_1,\ldots,s_J$.
We call this measure the local duration ratio (LDR).
The greater this ratio, the longer the duration of the converted utterance is 
relative to the reference utterance.
In the following, we use 
the mean absolute difference between the LDRs and 1 (in percentages)
as the overall measure for the LDRs.  
Thus, the closer it is to zero, the better the performance.
For example,
if the converted speech is 2 times faster than the reference speech,
the LDR will be 0.5 everywhere, and so its mean absolute difference from 1 will be $50\%$.

\subsection{Baseline Methods}

\subsubsection{sprocket}
We chose the open-source VC system 
called sprocket \cite{Kobayashi2018short} for comparison in our experiments. 
To run this method,
we used the source code provided by its author \cite{Kobayashi2018url}.
Note that this system was used as a baseline system in the
Voice Conversion Challenge (VCC) 2018 \cite{Lorenzo-Trueba2018short}.

\subsubsection{RNN-S2S-VC}
To evaluate the effect of the fully convolutional architecture adopted in ConvS2S-VC, 
we implemented its recurrent counterpart \cite{Tanaka2019short}, inspired by the architecture introduced 
in a S2S model-based TTS system called Tacotron \cite{Wang2017short} 
and considered it as another baseline. 
Although the original Tacotron used mel-spectra as the acoustic features,
the baseline system was designed to use the same acoustic features as our system.
The architecture was specifically designed as follows.
The encoder consisted of a bottleneck fully-connected prenet followed by a stack of 
$1\times 1$ 1D GLU convolutions and a bi-directional LSTM layer. 
The decoder was an autoregressive content-based attention network, 
consisting of a  bottleneck fully-connected prenet followed by 
a stateful LSTM  layer producing the attention query, which was then passed 
to a stack of two uni-directional residual LSTM layers, followed by a linear projection to generate the features. 
Note that we replaced all rectified linear unit (ReLU) activations with GLUs as with our model. 
We also designed and implemented a many-to-many extension of the above RNN-based model.

\subsection{Objective Evaluations}

\subsubsection{Effect of regularization}

\begin{figure*}[t!]
\centering
\begin{minipage}[t]{.243\linewidth}
  \centerline{\includegraphics[width=.99\linewidth]{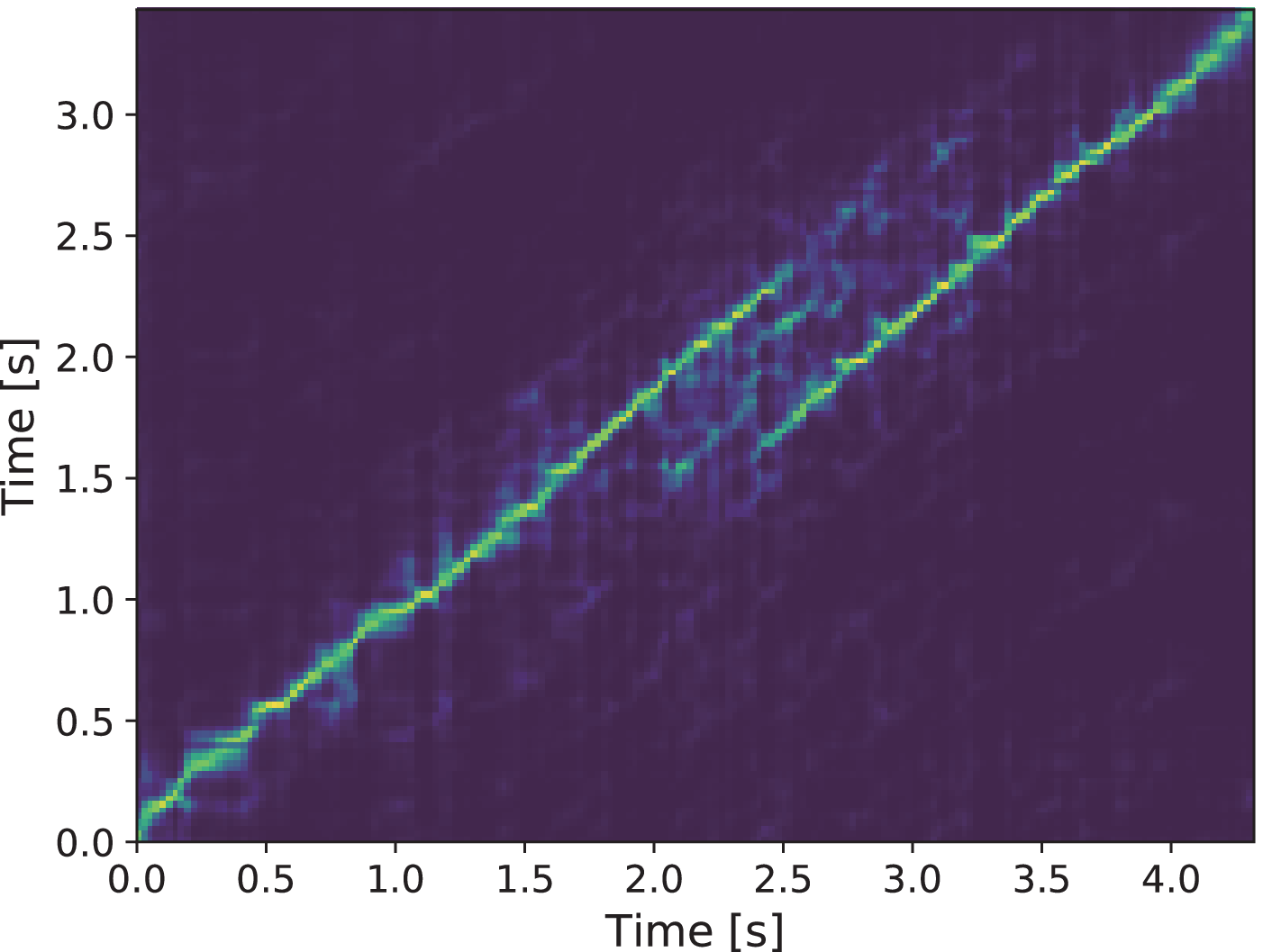}}
\end{minipage}
\begin{minipage}[t]{.243\linewidth}
  \centerline{\includegraphics[width=.99\linewidth]{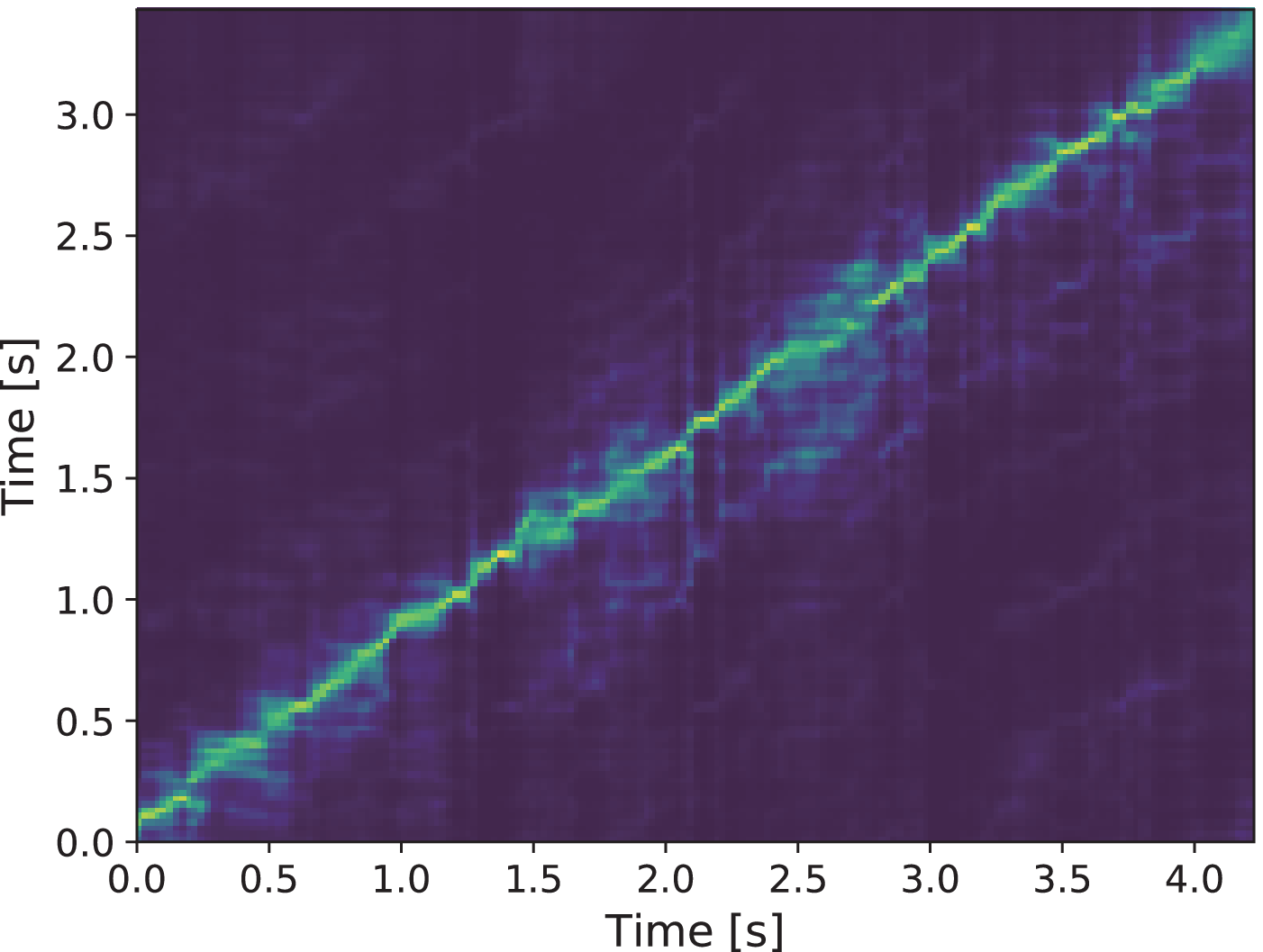}}
\end{minipage}
\begin{minipage}[t]{.243\linewidth}
  \centerline{\includegraphics[width=.99\linewidth]{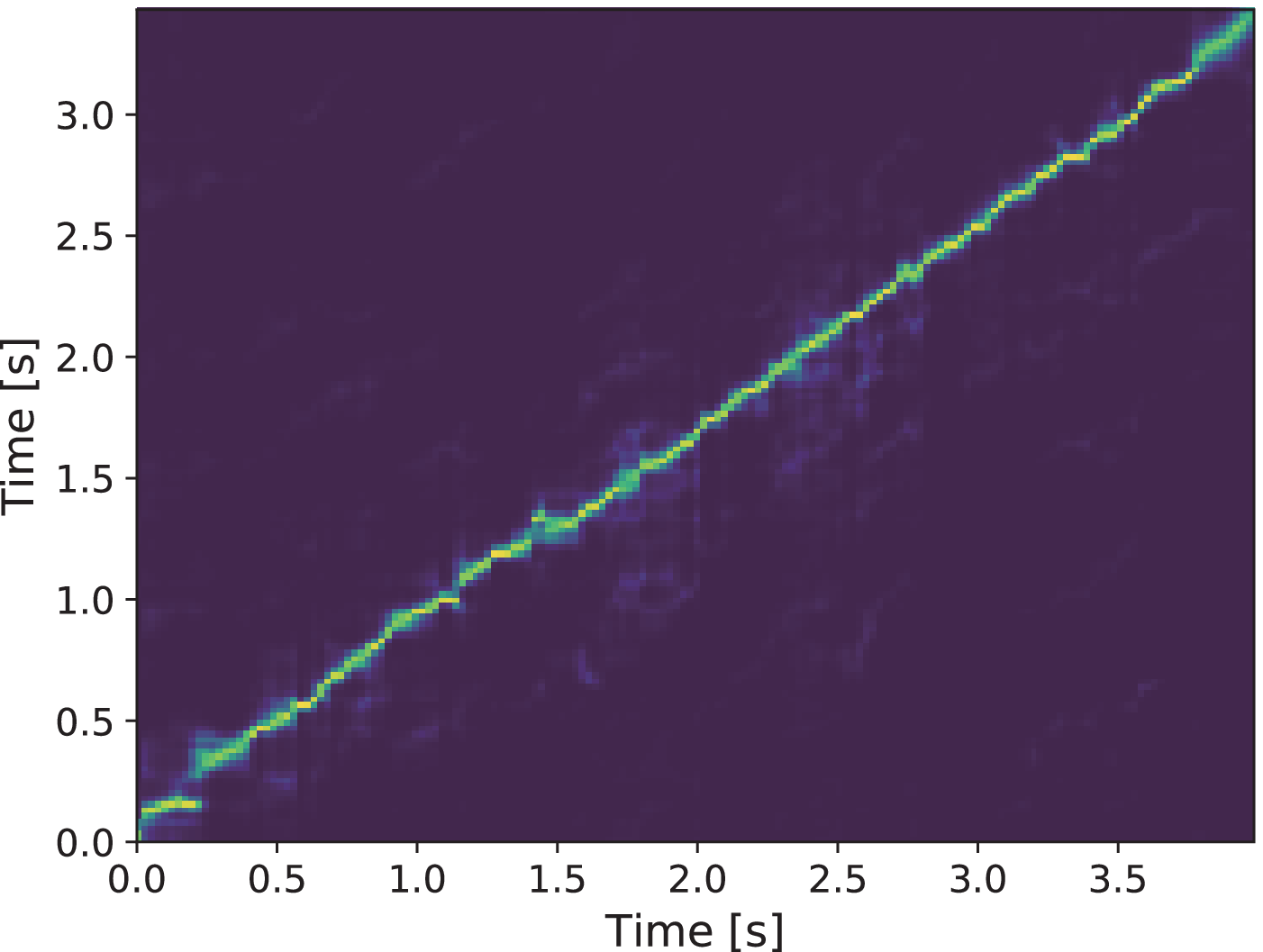}}
\end{minipage}
\begin{minipage}[t]{.243\linewidth}
  \centerline{\includegraphics[width=.99\linewidth]{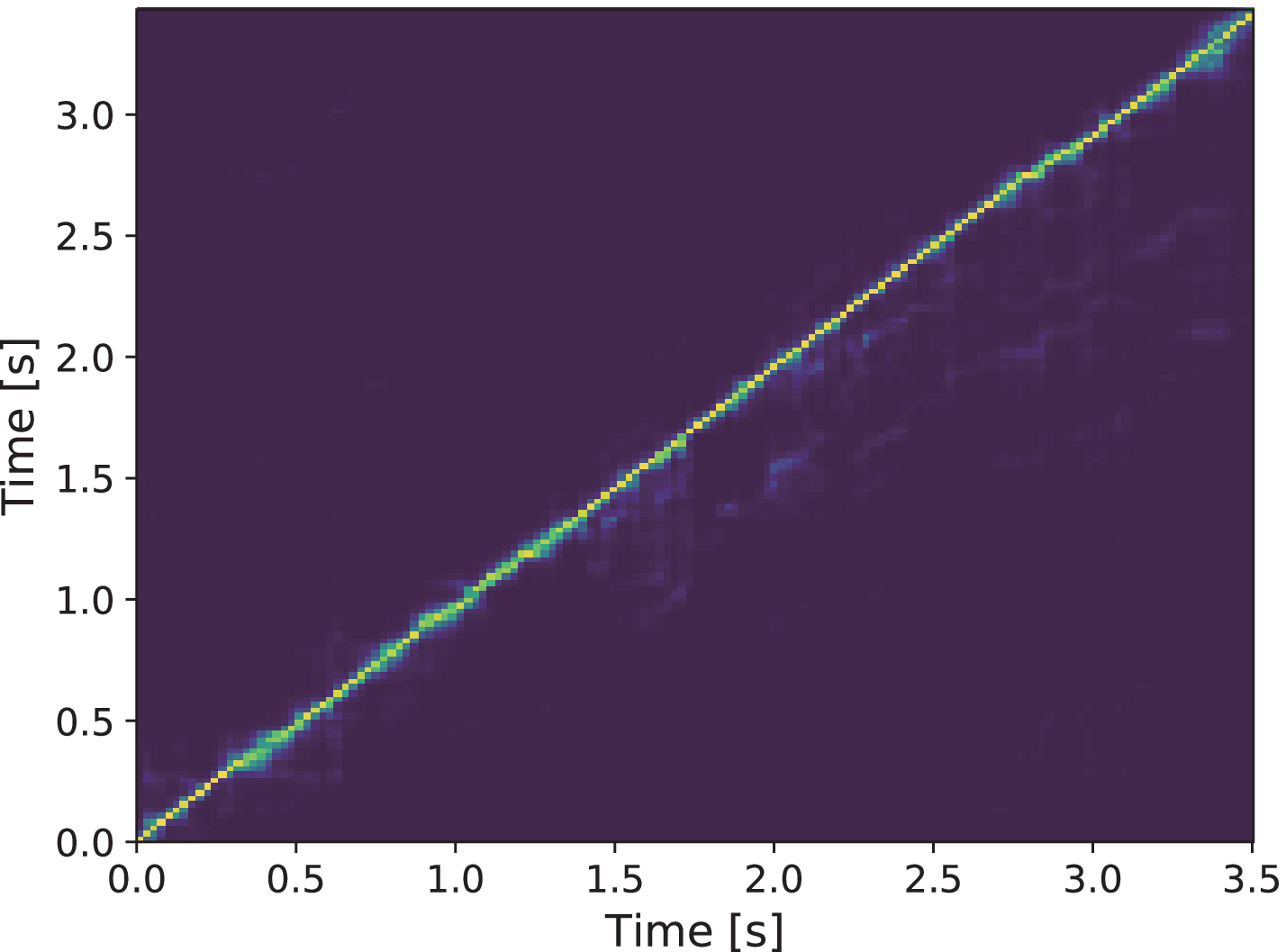}}
\end{minipage}
\vspace{-1ex}
\caption{Examples of the attention matrices predicted from 
test input female speech using the many-to-many model 
trained under four settings of $(\lambda_{\mathsf{r}},\lambda_{\mathsf{o}},\lambda_{\mathsf{i}})$:
$(0,0,0)$, 
$(1,0,0)$,
$(1,2000,0)$,
and
$(1,2000,1)$ (from left to right).
}
\label{fig:multidomain_reg_comp}
\end{figure*}

First, we evaluated the individual effects of 
the regularization techniques presented in
Subsections
\ref{subsec:model} and \ref{subsec:AL}
on both the pairwise and many-to-many models.
Tabs. \ref{tab:mcd_reg_comp}, \ref{tab:lfc_reg_comp}, and \ref{tab:ldr_reg_comp}
show the average MCDs (with $95\%$ confidence intervals), 
LFCs, and LDR deviations  
of the converted speech obtained 
using the pairwise and many-to-many models 
under 
different 
$(\lambda_{\mathsf{r}}, \lambda_{\mathsf{o}})$ settings 
$(0,0)$, 
$(1,0)$, 
and
$(1,2000)$
for the pairwise conversion model
and
different 
$(\lambda_{\mathsf{r}}, \lambda_{\mathsf{i}}, \lambda_{\mathsf{o}})$ settings
$(0,0,0)$, 
$(1,0,0)$, 
$(1,1,0)$,
and
$(1,1,2000)$
for the many-to-many model.
Owing to the limited amount of training data, 
the models trained without DAL did not successfully 
produce recognizable speech. 
Thus, we omit the results obtained when $\lambda_{\mathsf d}=0$.
As the results show, 
although there are a few exceptions, 
both the pairwise and many-to-many models performed better for most speaker pairs
in terms of the MCD measure
when 
all the regularization terms 
were simultaneously taken into account during training.
We also found that 
the effects of 
$L_{\mathsf{rec}}$ and $L_{\mathsf{oal}}$ on the LFC and LDR measures 
were less significant than on the MCD measure.
\reffig{multidomain_reg_comp} shows examples of
how each of the regularization techniques can affect the prediction of
the attention matrices by the many-to-many model at test time.
As these examples show, 
the CPL tended to have a notable effect on promoting
monotonicity and continuity of the attention prediction. 
However, it also had a negative effect of blurring the predicted attention distributions. 
The OAL and IML contributed to counteracting this negative effect by sharpening the attention matrices 
while keeping them monotonic and continuous.

\begin{table*}[t!]
\caption{Average MCDs [dB] 
obtained with the pairwise and many-to-many models 
trained with and without regularization.}
\centering
\begin{tabular}{l | l V{3} c|c|c|c|c|c|c}
\thline
\multicolumn{2}{c V{3}}{Speakers}&\multicolumn{3}{c|}{Pairwise}&\multicolumn{4}{c}{many-to-many}\\\hline
source&target
&$\lambda_{\mathsf r}\!=\!\lambda_{\mathsf o}\!=\!0$
&$\lambda_{\mathsf o}\!=\!0$
&full version
&$\lambda_{\mathsf r}\!=\!\lambda_{\mathsf i}\!=\!\lambda_{\mathsf o}\!=\!0$
&$\lambda_{\mathsf i}\!=\!\lambda_{\rm o}\!=\!0$
&$\lambda_{\rm o}\!=\!0$
&full version\\\thline
      &   bdl
      &$7.00\pm .09$&${\bf 6.84\pm .09}$&$6.93\pm .10$
      &$6.98\pm .10$&$6.76\pm .12$&$6.67\pm .14$&${\bf 6.57\pm .12}$\\
   clb&   slt
      &$6.37\pm .09$&${\bf 6.34\pm .10}$&$6.37\pm .08$
      &$6.34\pm .08$&$6.11\pm .05$&$6.01\pm .09$&${\bf 5.98\pm .08}$\\
      &   rms
      &$6.54\pm .10$&${\bf 6.52\pm .13}$&$6.53\pm .16$
      &$6.27\pm .05$&$6.23\pm .06$&$6.44\pm .12$&${\bf 6.11\pm .08}$\\\hline
      &   clb
      &$6.30\pm .06$&$6.09\pm .08$&${\bf 6.03\pm .09}$
      &$6.03\pm .07$&$6.05\pm .11$&$5.91\pm .08$&${\bf 5.86\pm .09}$\\
   bdl&   slt
      &$6.61\pm .10$&$6.66\pm .10$&${\bf 6.51\pm .12}$
      &$6.58\pm .12$&$6.47\pm .06$&$6.25\pm .09$&${\bf 6.19\pm .11}$\\
      &   rms
      &$6.75\pm .11$&${\bf 6.68\pm .13}$&$6.79\pm .16$
      &$6.65\pm .14$&$6.45\pm .07$&$6.44\pm .11$&${\bf 6.24\pm .12}$\\\hline
      &   clb
      &$6.21\pm .12$&$6.12\pm .08$&${\bf 6.03\pm .06}$
      &$6.08\pm .07$&$6.14\pm .09$&$5.79\pm .09$&${\bf 5.78\pm .16}$\\
   slt&   bdl
      &$7.25\pm .16$&$7.21\pm .18$&${\bf 7.07\pm .16}$
      &$7.10\pm .14$&$6.99\pm .14$&$6.80\pm .14$&${\bf 6.72\pm .14}$\\
      &   rms
      &$6.61\pm .07$&${\bf 6.56\pm .08}$&$6.61\pm .09$
      &$6.50\pm .09$&$6.44\pm .12$&$6.51\pm .10$&${\bf 6.27\pm .08}$\\\hline
      &   clb
      &$6.42\pm .14$&$6.37\pm .18$&${\bf 6.30\pm .11}$
      &$6.05\pm .06$&$6.10\pm .11$&$5.96\pm .09$&${\bf 5.93\pm .11}$\\
   rms&   bdl
      &$7.16\pm .10$&$7.14\pm .14$&${\bf 7.08\pm .15}$
      &$7.07\pm .10$&$6.91\pm .11$&$6.74\pm .11$&${\bf 6.67\pm .12}$\\
      &   slt
      &$6.78\pm .19$&$6.72\pm .23$&${\bf 6.50\pm .07}$
      &$6.49\pm .11$&${\bf 6.28\pm .11}$&$6.35\pm .16$&$6.32\pm .16$\\\hline 
\multicolumn{2}{c V{3}}{All pairs}
      &$6.67\pm .04$&$6.61\pm .05$&${\bf 6.56\pm .05}$
      &$6.51\pm .04$&$6.41\pm .04$&$6.32\pm .04$&${\bf 6.22\pm .04}$\\\thline
\end{tabular}
\label{tab:mcd_reg_comp}
\end{table*}

\begin{table*}[t!]
\caption{Average LFCs obtained with the pairwise and many-to-many models  
trained with and without regularization.}
\centering
\begin{tabular}{l | l V{3} c|c|c|c|c|c|c}
\thline
\multicolumn{2}{c V{3}}{Speakers}&\multicolumn{3}{c|}{Pairwise}&\multicolumn{4}{c}{many-to-many}\\\hline
source&target
&$\lambda_{\rm r}\!=\!\lambda_{\rm o}\!=\!0$
&$\lambda_{\rm o}\!=\!0$
&full version
&$\lambda_{\rm r}\!=\!\lambda_{\rm i}\!=\!\lambda_{\rm o}\!=\!0$
&$\lambda_{\rm i}\!=\!\lambda_{\rm o}\!=\!0$
&$\lambda_{\rm o}\!=\!0$
&full version
\\\thline
      &   bdl
      &$0.852$&$0.856$&${\bf 0.869}$
      &${\bf 0.876}$&$0.874$&$0.851$&$0.845$\\
   clb&   slt
      &${\bf 0.846}$&$0.835$&$0.833$
      &$0.834$&${\bf 0.852}$&$0.831$&$0.841$\\
      &   rms
      &${\bf 0.829}$&$0.805$&$0.771$
      &${\bf 0.835}$&$0.741$&$0.751$&$0.811$\\\hline
      &   clb
      &${\bf 0.844}$&$0.815$&$0.810$
      &${\bf 0.846}$&$0.835$&$0.831$&$0.823$\\
   bdl&   slt
      &$0.768$&$0.799$&${\bf 0.815}$
      &$0.775$&$0.792$&${\bf 0.875}$&$0.864$\\
      &   rms
      &${\bf 0.805}$&$0.750$&$0.800$
      &$0.759$&$0.742$&$0.788$&${\bf 0.855}$\\\hline
      &   clb
      &$0.838$&$0.796$&${\bf 0.861}$
      &$0.795$&$0.770$&${\bf 0.830}$&$0.812$\\
   slt&   bdl
      &$0.821$&$0.832$&${\bf 0.850}$
      &$0.860$&$0.859$&${\bf 0.861}$&$0.850$\\
      &   rms
      &${\bf 0.804}$&$0.797$&$0.785$
      &$0.789$&$0.783$&$0.759$&${\bf 0.799}$\\\hline
      &   clb
      &${\bf 0.850}$&$0.830$&$0.821$
      &$0.833$&$0.794$&${\bf 0.834}$&$0.819$\\
   rms&   bdl
      &${\bf 0.854}$&$0.853$&$0.825$
      &$0.864$&$0.845$&${\bf 0.877}$&$0.870$\\
      &   slt
      &$0.792$&${\bf 0.825}$&$0.809$
      &$0.776$&$0.794$&${\bf 0.866}$&$0.826$\\\hline
\multicolumn{2}{c V{3}}{All pairs}
      &${\bf 0.829}$&$0.818$&$0.826$
      &$0.833$&$0.822$&$0.834$&${\bf 0.838}$\\\thline
\end{tabular}
\label{tab:lfc_reg_comp}
\end{table*}

\begin{table*}[t!]
\caption{Average LDR deviations (\%) obtained with the pairwise and many-to-many models  
trained with and without regularization.}
\centering
\begin{tabular}{l | l V{3} c|c|c|c|c|c|c}
\thline
\multicolumn{2}{c V{3}}{Speakers}&\multicolumn{3}{c|}{Pairwise}&\multicolumn{4}{c}{many-to-many}\\\hline
source&target&$\lambda_{\rm r}\!=\!\lambda_{\rm o}\!=\!0$
&$\lambda_{\rm o}\!=\!0$
&full version
&$\lambda_{\mathsf r}\!=\!\lambda_{\mathsf i}\!=\!\lambda_{\mathsf o}\!=\!0$
&$\lambda_{\mathsf i}\!=\!\lambda_{\mathsf o}\!=\!0$
&$\lambda_{\mathsf o}\!=\!0$
&full version\\\thline
      &   bdl&$5.28$&${\bf 3.54}$&$5.16$&$3.96$&${\bf 3.48}$&$6.14$&$8.58$\\
   clb&   slt&$2.42$&$3.76$&${\bf 1.96}$&$2.61$&$3.43$&${\bf 0.55}$&$5.85$\\
      &   rms&${\bf 2.59}$&$4.25$&$5.86$&${\bf 2.34}$&$10.44$&$3.01$&$4.20$\\\hline
      &   clb&$4.96$&$3.87$&${\bf 1.97}$&$5.66$&$5.20$&$3.78$&${\bf 3.47}$\\
   bdl&   slt&$6.16$&${\bf 4.30}$&$6.78$&$6.43$&$4.09$&${\bf 3.56}$&$5.53$\\
      &   rms&$4.22$&$6.41$&${\bf 0.79}$&${\bf 2.45}$&$5.43$&$5.25$&$6.06$\\\hline
      &   clb&$2.59$&${\bf 0.60}$&$0.81$&${\bf 0.31}$&$0.75$&$2.45$&$0.66$\\
   slt&   bdl&$7.36$&${\bf 5.59}$&$6.70$&${\bf 1.73}$&$3.64$&$4.25$&$5.49$\\
      &   rms&$4.49$&$4.30$&${\bf 4.08}$&$4.17$&$11.40$&${\bf 2.76}$&$4.66$\\\hline
      &   clb&${\bf 1.04}$&$3.07$&$1.99$&$5.40$&$4.83$&${\bf 2.94}$&$3.55$\\
   rms&   bdl&${\bf 2.87}$&$5.10$&$6.87$&${\bf 2.59}$&$3.63$&$5.13$&$10.85$\\
      &   slt&${\bf 2.88}$&$7.16$&$3.29$&$6.93$&$4.62$&$3.38$&${\bf 2.77}$\\\hline
 \multicolumn{2}{c V{3}}{All pairs}
      &$4.17$&$4.21$&${\bf 3.47}$
      &${\bf 3.51}$&$4.48$&$4.01$&$4.65$\\\thline
\end{tabular}
\label{tab:ldr_reg_comp}
\end{table*}

\subsubsection{Comparison of normalization methods}

\begin{table*}[t!]
\caption{MCD [dB] Comparison of normalization methods. 
}
\centering
\begin{tabular}{l | l V{3} c|c|c|c|c|c|c|c}
\thline
\multicolumn{2}{c V{3}}{Speakers}&\multicolumn{3}{c|}{Pairwise}&\multicolumn{5}{c}{many-to-many}\\\hline
source&target
&IN
&WN
&BN
&IN
&CIN
&WN
&BN
&CBN\\\thline
      &   bdl
      &$7.76\pm .19$&$7.10\pm .11$&${\bf 6.93\pm .10}$
      &$9.04\pm .21$&$9.00\pm .25$&$6.83\pm .13$&$7.42\pm .10$&${\bf 6.57\pm .12}$\\
   clb&   slt
      &$6.79\pm .10$&$6.63\pm .10$&${\bf 6.37\pm .08}$
      &$8.16\pm .15$&$7.96\pm .17$&$6.16\pm .07$&$6.96\pm .09$&${\bf 5.98\pm .08}$\\
      &   rms
      &$7.57\pm .25$&$6.71\pm .12$&${\bf 6.53\pm .16}$
      &$7.72\pm .23$&$7.96\pm .21$&$6.20\pm .08$&$8.01\pm .05$&${\bf 6.11\pm .08}$\\\hline
      &   clb
      &$6.85\pm .12$&$6.38\pm .07$&${\bf 6.03\pm .09}$
      &$7.93\pm .26$&$7.52\pm .30$&${\bf 5.85\pm .07}$&$8.04\pm .20$&$5.86\pm .09$\\
   bdl&   slt
      &$7.32\pm .20$&$6.69\pm .06$&${\bf 6.51\pm .12}$
      &$7.97\pm .18$&$7.90\pm .21$&$6.36\pm .08$&$8.03\pm .25$&${\bf 6.19\pm .11}$\\
      &   rms
      &$7.49\pm .16$&$6.93\pm .12$&${\bf 6.79\pm .16}$
      &$7.86\pm .18$&$7.69\pm .24$&$6.31\pm .09$&$8.06\pm .13$&${\bf 6.24\pm .12}$\\\hline
      &   clb
      &$6.64\pm .15$&$6.24\pm .05$&${\bf 6.03\pm .06}$
      &$7.68\pm .28$&$8.26\pm .26$&$5.82\pm .09$&$8.06\pm .21$&${\bf 5.78\pm .16}$\\
   slt&   bdl
      &$7.95\pm .24$&$7.28\pm .16$&${\bf 7.07\pm .16}$
      &$9.14\pm .23$&$9.61\pm .26$&$6.95\pm .13$&$7.73\pm .20$&${\bf 6.72\pm .14}$\\
      &   rms
      &$7.32\pm .19$&$6.76\pm .12$&${\bf 6.61\pm .09}$
      &$7.60\pm .14$&$7.69\pm .14$&$6.42\pm .14$&$8.54\pm .09$&${\bf 6.27\pm .08}$\\\hline
      &   clb
      &$6.96\pm .15$&$6.43\pm .12$&${\bf 6.30\pm .11}$
      &$7.69\pm .20$&$7.99\pm .26$&$6.04\pm .13$&$7.93\pm .20$&${\bf 5.93\pm .11}$\\
   rms&   bdl
      &$8.04\pm .19$&$7.25\pm .12$&${\bf 7.08\pm .15}$
      &$8.97\pm .26$&$9.16\pm .28$&$6.93\pm .11$&$7.88\pm .39$&${\bf 6.67\pm .12}$\\
      &   slt
      &$7.48\pm .34$&$6.92\pm .12$&${\bf 6.50\pm .07}$
      &$8.32\pm .21$&$8.15\pm .25$&$6.47\pm .17$&$8.43\pm .62$&${\bf 6.32\pm .16}$\\\hline 
\multicolumn{2}{c V{3}}{All pairs}
      &$7.34\pm .07$&$6.78\pm .05$&${\bf 6.56\pm .05}$
      &$8.17\pm .08$&$8.24\pm .90$&$6.36\pm .05$&$7.92\pm .08$&${\bf 6.22\pm .04}$\\\thline
\end{tabular}
\label{tab:mcd_norm_comp}
\end{table*}

\begin{table*}[t!]
\caption{LFC comparison of normalization methods.}
\centering
\begin{tabular}{l | l V{3} c|c|c|c|c|c|c|c}
\thline
\multicolumn{2}{c V{3}}{Speakers}&\multicolumn{3}{c|}{Pairwise}&\multicolumn{5}{c}{many-to-many}\\\hline
source&target
&IN
&WN
&BN
&IN
&CIN
&WN
&BN
&CBN
\\\thline
      &   bdl
      &$0.827$&$0.840$&${\bf 0.869}$
      &$0.556$&$0.456$&${\bf 0.873}$&$0.769$&$0.845$\\
   clb&   slt
      &$0.819$&$0.813$&${\bf 0.833}$
      &$0.547$&$0.794$&$0.830$&$0.815$&${\bf 0.841}$\\
      &   rms
      &$0.657$&$0.770$&${\bf 0.771}$
      &$0.475$&$0.338$&${\bf 0.831}$&$0.494$&$0.811$\\\hline
      &   clb
      &$0.781$&$0.779$&${\bf 0.810}$
      &$0.766$&$0.725$&${\bf 0.829}$&$0.652$&$0.823$\\
   bdl&   slt
      &$0.746$&${\bf 0.831}$&$0.815$
      &$0.787$&$0.693$&$0.789$&$0.617$&${\bf 0.864}$\\
      &   rms
      &$0.634$&$0.738$&${\bf 0.800}$
      &$0.588$&$0.608$&$0.763$&$0.444$&${\bf 0.855}$\\\hline
      &   clb
      &$0.811$&$0.835$&${\bf 0.861}$
      &$0.594$&$0.647$&$0.803$&$0.745$&${\bf 0.812}$\\
   slt&   bdl
      &$0.792$&$0.819$&${\bf 0.850}$
      &$0.537$&$0.356$&${\bf 0.852}$&$0.748$&${\bf 0.850}$\\
      &   rms
      &$0.680$&$0.729$&${\bf 0.785}$
      &$0.504$&$0.327$&$0.783$&$0.293$&${\bf 0.799}$\\\hline
      &   clb
      &$0.738$&$0.761$&${\bf 0.821}$
      &$0.530$&$0.471$&$0.794$&$0.723$&${\bf 0.819}$\\
   rms&   bdl
      &${\bf 0.841}$&$0.787$&$0.825$
      &$0.675$&$0.479$&$0.853$&$0.709$&${\bf 0.870}$\\
      &   slt
      &$0.678$&$0.768$&${\bf 0.809}$
      &$0.525$&$0.657$&$0.781$&$0.554$&${\bf 0.826}$\\\hline
\multicolumn{2}{c V{3}}{All pairs}
      &$0.772$&$0.800$&${\bf 0.826}$
      &$0.582$&$0.572$&$0.818$&$0.672$&${\bf 0.838}$\\\thline
\end{tabular}
\label{tab:lfc_norm_comp}
\end{table*}

\begin{table*}[t!]
\caption{LDR deviation (\%) comparison of normalization methods.}
\centering
\begin{tabular}{l | l V{3} c|c|c|c|c|c|c|c}
\thline
\multicolumn{2}{c V{3}}{Speakers}&\multicolumn{3}{c|}{Pairwise}&\multicolumn{5}{c}{many-to-many}\\\hline
source&target
&IN
&WN
&BN
&IN
&CIN
&WN
&BN
&CBN\\\thline
      &   bdl&${\bf 2.10}$&$5.72$&$5.16$&$9.69$&$15.10$&${\bf 4.76}$&$6.91$&$8.58$\\
   clb&   slt&$3.46$&${\bf 1.36}$&$1.96$&$13.38$&$13.56$&${\bf 2.04}$&$3.68$&$5.85$\\
      &   rms&$4.20$&${\bf 2.69}$&$5.86$&$15.22$&$23.06$&$6.01$&$5.62$&${\bf 4.20}$\\\hline
      &   clb&$4.61$&$4.36$&${\bf 1.97}$&$15.59$&$25.97$&$4.83$&$7.91$&${\bf 3.47}$\\
   bdl&   slt&$9.36$&${\bf 3.72}$&$6.78$&$15.49$&$13.45$&$4.14$&${\bf 2.24}$&$5.53$\\
      &   rms&$3.09$&$2.62$&${\bf 0.79}$&$19.81$&$26.39$&${\bf 5.37}$&$8.64$&$6.06$\\\hline
      &   clb&$2.01$&$1.52$&${\bf 0.81}$&$10.67$&$15.91$&$4.13$&$3.46$&${\bf 0.66}$\\
   slt&   bdl&${\bf 3.91}$&$8.31$&$6.70$&$16.96$&$19.57$&${\bf 3.12}$&$7.78$&$5.49$\\
      &   rms&$6.16$&$4.14$&${\bf 4.08}$&$18.91$&$24.02$&$4.68$&${\bf 0.01}$&$4.66$\\\hline
      &   clb&$2.22$&$2.51$&${\bf 1.99}$&$14.70$&$20.74$&${\bf 2.48}$&$7.44$&$3.55$\\
   rms&   bdl&$5.23$&${\bf 4.38}$&$6.87$&$16.41$&$13.97$&${\bf 5.26}$&$9.68$&$10.85$\\
      &   slt&${\bf 3.25}$&$5.35$&$3.29$&$10.76$&$13.63$&$4.31$&$5.23$&${\bf 2.77}$\\\hline
 \multicolumn{2}{c V{3}}{All pairs}
      &$3.87$&$3.87$&${\bf 3.47}$
      &$13.77$&$18.03$&${\bf 4.46}$&$4.77$&$4.65$\\\thline
\end{tabular}
\label{tab:ldr_norm_comp}
\end{table*}

For normalization, 
there are several choices 
including 
instance normalization (IN) \cite{Ulyanov2016short},
weight normalization (WN) \cite{Salimans2016short},
and
batch normalization (BN) \cite{Ioffe2015short}.
For our many-to-many model,
other choices include 
conditional IN (CIN) and 
conditional BN (CBN).
We compared the effects of these normalization methods on both
the pairwise and many-to-many models on the basis of the 
MCD, LFC, and LDR measures.
Note that all the normalization layers in 
Tab. \ref{tab:architectures} are 
excluded in the WN counterparts.
The average MCDs, LFCs, and LDR deviations obtained using
these normalization methods 
are demonstrated in 
Tabs. \ref{tab:mcd_norm_comp}, \ref{tab:lfc_norm_comp}, and \ref{tab:ldr_norm_comp}.
As the results show, BN worked better than IN and WN when applied to the pairwise conversion model
especially in terms of the MCD and LFC measures.
However, naively applying it directly to the 
many-to-many model did not work satisfactorily,
as expected in \refsubsec{dbn}.
This was also the case with IN. 
Although CIN was found to perform poorly, 
CBN worked significantly better.

\begin{table*}[t!]
\centering
\caption{Average MCDs (dB) with 95\% confidence intervals
obtained with the baseline and proposed methods}
\begin{tabular}{l | l V{3} c|c|c|c|c}
\thline
\multicolumn{2}{c V{3}}{Speakers}&\multirow{2}{*}{sprocket}&\multicolumn{2}{c|}{RNN-S2S}&\multicolumn{2}{c}{proposed (ConvS2S)}\\\cline{1-2}\cline{4-7}
source&target&&pairwise&many-to-many&pairwise &many-to-many\\\thline
      &   bdl&$6.98\pm .10$&$6.87\pm .09$&$6.94\pm .15$&$6.93\pm .10$&${\bf 6.57\pm .12}$\\
   clb&   slt&$6.34\pm .06$&$6.22\pm .07$&$6.26\pm .09$&$6.37\pm .08$&${\bf 5.98\pm .08}$\\
      &   rms&$6.84\pm .07$&$6.45\pm .09$&$6.23\pm .06$&$6.53\pm .16$&${\bf 6.11\pm .08}$\\\hline
      &   clb&$6.44\pm .10$&$6.21\pm .13$&$6.02\pm .12$&$6.03\pm .09$&${\bf 5.86\pm .09}$\\
   bdl&   slt&$6.46\pm .04$&$6.68\pm .11$&$6.38\pm .14$&$6.51\pm .12$&${\bf 6.19\pm .11}$\\
      &   rms&$7.24\pm .12$&$6.69\pm .21$&$6.35\pm .09$&$6.79\pm .16$&${\bf 6.24\pm .12}$\\\hline
      &   clb&$6.21\pm .06$&$6.13\pm .10$&$6.03\pm .10$&$6.03\pm .06$&${\bf 5.78\pm .16}$\\
   slt&   bdl&$6.80\pm .05$&$7.08\pm .11$&$7.09\pm .12$&$7.07\pm .16$&${\bf 6.72\pm .14}$\\
      &   rms&$6.87\pm .10$&$6.64\pm .13$&$6.38\pm .07$&$6.61\pm .09$&${\bf 6.27\pm .08}$\\\hline
      &   clb&$6.43\pm .06$&$6.26\pm .14$&$6.23\pm .12$&$6.30\pm .11$&${\bf 5.93\pm .11}$\\
   rms&   bdl&$7.40\pm .15$&$7.11\pm .16$&$7.22\pm .16$&$7.08\pm .15$&${\bf 6.67\pm .12}$\\
      &   slt&$6.76\pm .09$&$6.53\pm .11$&$6.41\pm .12$&$6.50\pm .07$&${\bf 6.32\pm .16}$\\\hline
\multicolumn{2}{c V{3}}{All pairs}
             &$6.73\pm .03$&$6.57\pm .05$&$6.46\pm .05$&$6.56\pm .05$&${\bf 6.22\pm .04}$\\\thline
\end{tabular}
\label{tab:mcd_baseline_comp}
\end{table*}

\begin{table*}[t!]
\centering
\caption{LFCs obtained with the baseline and proposed methods}
\begin{tabular}{l | l V{3} c|c|c|c|c}
\thline
\multicolumn{2}{c V{3}}{Speakers}&\multirow{2}{*}{sprocket}&\multicolumn{2}{c|}{RNN-S2S}&\multicolumn{2}{c}{proposed (ConvS2S)}\\\cline{1-2}\cline{4-7}
source&target&&pairwise&many-to-many&pairwise &many-to-many\\\thline
      &   bdl&$0.643$&$0.851$&${\bf 0.875}$&$0.869$&$0.847$\\
   clb&   slt&$0.790$&$0.765$&$0.815$&$0.833$&${\bf 0.845}$\\
      &   rms&$0.556$&$0.784$&$0.787$&$0.771$&${\bf 0.795}$\\\hline
      &   clb&$0.642$&$0.748$&${\bf 0.840}$&$0.810$&$0.826$\\
   bdl&   slt&$0.632$&$0.738$&$0.797$&$0.815$&${\bf 0.863}$\\
      &   rms&$0.467$&$0.719$&$0.715$&$0.800$&${\bf 0.829}$\\\hline
      &   clb&$0.820$&$0.847$&$0.776$&${\bf 0.861}$&$0.827$\\
   slt&   bdl&$0.663$&$0.812$&$0.834$&$0.850$&${\bf 0.852}$\\
      &   rms&$0.611$&$0.753$&$0.773$&$0.785$&${\bf 0.806}$\\\hline
      &   clb&$0.632$&$0.753$&$0.818$&${\bf 0.821}$&$0.796$\\
   rms&   bdl&$0.648$&$0.817$&$0.854$&$0.825$&${\bf 0.877}$\\
      &   slt&$0.674$&$0.783$&$0.785$&$0.809$&${\bf 0.838}$\\\hline
\multicolumn{2}{c V{3}}{All pairs}
      &$0.653$&$0.798$&$0.808$&$0.826$&${\bf 0.836}$\\\thline   
\end{tabular}
\label{tab:lfc_baseline_comp}
\end{table*}
\begin{table*}[t!]
\centering
\caption{LDR deviations (\%) obtained with the baseline and proposed methods}
\begin{tabular}{l | l V{3} c|c|c|c|c}
\thline
\multicolumn{2}{c V{3}}{Speakers}&\multirow{2}{*}{sprocket}&\multicolumn{2}{c|}{RNN-S2S}&\multicolumn{2}{c}{proposed (ConvS2S)}\\\cline{1-2}\cline{4-7}
source&target&&pairwise&many-to-many&pairwise &many-to-many\\\thline
      &   bdl&$17.66$&${\bf 0.52}$&$1.30$&$5.16$&$8.58$\\
   clb&   slt&$9.74$&$2.95$&${\bf 1.24}$&$1.96$&$5.85$\\
      &   rms&$3.24$&${\bf 2.27}$&$4.92$&$5.86$&$4.20$\\\hline
      &   clb&$16.65$&$3.52$&$4.94$&${\bf 1.97}$&$3.47$\\
   bdl&   slt&${\bf 4.58}$&$7.76$&$7.18$&$6.78$&$5.53$\\
      &   rms&$15.20$&$2.65$&$3.72$&${\bf 0.79}$&$6.06$\\\hline
      &   clb&$9.25$&$2.63$&$3.49$&${\bf 0.81}$&$0.66$\\
   slt&   bdl&$5.52$&$4.61$&${\bf 0.01}$&$6.70$&$5.49$\\
      &   rms&$11.46$&${\bf 3.36}$&$3.92$&$4.08$&$4.66$\\\hline
      &   clb&$2.84$&$2.80$&$5.40$&${\bf 1.99}$&$3.55$\\
   rms&   bdl&$17.76$&$4.53$&${\bf 3.19}$&$6.87$&$10.85$\\
      &   slt&$11.95$&$6.84$&$4.15$&$3.29$&${\bf 2.77}$\\\hline
\multicolumn{2}{c V{3}}{All pairs}
      &$10.60$&$3.62$&$3.56$&${\bf 3.47}$&$4.65$\\\thline  
\end{tabular}
\label{tab:ldr_baseline_comp}
\end{table*}

\subsubsection{Comparisons with baseline methods}

Tabs. \ref{tab:mcd_baseline_comp}, \ref{tab:lfc_baseline_comp}, and \ref{tab:ldr_baseline_comp} 
show the average MCDs, 
LFCs, and LDRs obtained with the proposed and baseline methods.
As 
Tabs. \ref{tab:mcd_baseline_comp} and \ref{tab:lfc_baseline_comp}
show, 
the pairwise versions of ConvS2S-VC and RNN-S2S-VC performed comparably to each other
and significantly better than sprocket.
The effect of the many-to-many extension was noticeable 
for both ConvS2S-VC and RNN-S2S-VC,
revealing the advantage of exploiting the training data of all the speakers.
The many-to-many ConvS2S-VC performed better than its RNN counterpart.
This demonstrates the effect of the convolutional architecture. 
Since 
sprocket is designed to 
keep the speaking rate and rhythm of input speech unchanged,
the performance gains over sprocket in terms of the LDR measure show
how well the competing methods are able to predict 
the speaking rate and rhythm of target speech.
As Tab. \ref{tab:ldr_baseline_comp} shows, 
both the pairwise and many-to-many versions of
RNN-S2S-VC and ConvS2S-VC obtained LDR deviations closer to 0 than sprocket. 

As mentioned earlier, 
one important advantage of the proposed model over its RNN counterpart 
is that it can be trained efficiently 
thanks to the nature of the convolutional architectures.
In fact, whereas  
the pairwise and many-to-many versions of the RNN-based model
took about 30 and 50 hours to train, 
the two versions of the proposed model only took about 4 and 7 hours to train 
under the current experimental settings.
We implemented all the algorithms in PyTorch
and used a single Tesla V100 GPU with a 32.0 GB memory
for training each model.

\begin{table}[t]
\caption{Average MCDs (dB), 
LFCs, and LDR deviations (\%) obtained with
the any-to-many setting 
under a closed-set condition.}
\centering
\begin{tabular}{l| l V{3} c|c|c}
\thline
\multicolumn{2}{c V{3}}{Speaker pair}&\multicolumn{3}{c}{Measures}\\\hline
source&target&MCD{\tiny (dB)}&LFC&LDR{\tiny (\%)}\\\thline
\multirow{3}{*}{clb} 
   &bdl&$6.81\pm .10$&$0.907$&$11.33$\\
   &slt&$6.27\pm .06$&$0.842$&$3.70$\\
   &rms&$6.26\pm .07$&$0.794$&$7.54$\\\hline
\multirow{3}{*}{bdl} 
   &clb&$6.10\pm .10$&$0.849$&$3.68$\\
   &slt&$6.55\pm .13$&$0.826$&$8.59$\\
   &rms&$6.49\pm .13$&$0.811$&$2.48$\\\hline
\multirow{3}{*}{slt} 
   &clb&$6.02\pm .09$&$0.813$&$1.90$\\
   &bdl&$7.03\pm .13$&$0.878$&$5.87$\\
   &rms&$6.44\pm .07$&$0.810$&$4.17$\\\hline
\multirow{3}{*}{rms} 
   &clb&$6.33\pm .13$&$0.815$&$4.73$\\
   &bdl&$6.95\pm .09$&$0.825$&$10.77$\\
   &slt&$6.51\pm .13$&$0.868$&$3.62$\\\hline
\multicolumn{2}{c V{3}}{All pairs}&$6.48\pm .04$&$0.829$&$4.67$\\\thline
\end{tabular}
\label{tab:all_m2o_seen}
\end{table}

\begin{table}[t]
\centering
\caption{Average MCDs (dB), LFCs, and LDR deviations (\%) obtained with
the any-to-many setting 
under an open-set condition.}
\begin{tabular}{l| l V{3} c|c|c}
\thline
\multicolumn{2}{c V{3}}{Speaker pair}&\multicolumn{3}{c}{Measures}\\\hline
source&target&MCD{\tiny (dB)}&LFC&LDR{\tiny (\%)}\\\thline
\multirow{4}{*}{lnh} 
   &clb&$6.29\pm .12$&$0.778$&$2.74$\\
   &bdl&$7.12\pm .12$&$0.803$&$9.06$\\
   &slt&$6.44\pm .07$&$0.694$&$0.45$\\
   &rms&$6.67\pm .09$&$0.720$&$4.44$\\\hline
\multicolumn{2}{c V{3}}{All pairs}&$6.63\pm .07$&$0.747$&$4.36$\\\thline
\end{tabular}
\label{tab:all_m2o_unseen}
\end{table}

\begin{table}[t]
\centering
\caption{Average MCDs (dB), LFCs, and LDR deviations (\%) obtained with
sprocket under a speaker-dependent condition.}
\begin{tabular}{l| l V{3} c|c|c}
\thline
\multicolumn{2}{c V{3}}{Speaker pair}&\multicolumn{3}{c}{Measures}\\\hline
source&target&MCD{\tiny (dB)}&LFC&LDR{\tiny \%}\\\thline
\multirow{4}{*}{lnh} 
   &clb&$6.76\pm .08$&$0.716$&$6.61$\\
   &bdl&$8.26\pm .35$&$0.523$&$13.38$\\
   &slt&$6.62\pm .11$&$0.771$&$5.72$\\
   &rms&$7.22\pm .10$&$0.480$&$4.87$\\\hline
\multicolumn{2}{c V{3}}{All pairs}&$7.21\pm .14$&$0.579$&$7.61$\\\thline
\end{tabular}
\label{tab:sprocket_lnh}
\end{table}

\subsubsection{Performance of any-to-many setting}

The modifications described in \refsubsec{many2one} 
make it possible to handle 
any-to-many VC tasks.
We evaluated how these modifications actually affected the performance.
Tab. \ref{tab:all_m2o_seen}
shows the average MCDs, LFCs, and LDR deviations obtained with 
the any-to-many setting
under 
a closed-set condition, 
where 
the speaker of input speech is unknown
but is seen in the training data. 
Whereas the pairwise and 
the default many-to-many versions
must be informed about
the speaker of each input utterance at test time, 
the any-to-many version requires no information. 
This can be convenient in practical scenarios of VC applications, 
but because of the disadvantage in the test condition,
the problem becomes more challenging.
As the results show, 
the MCDs and LFCs obtained with the any-to-many version
were only slightly worse than those obtained with the 
default many-to-many model
despite this disadvantage.
It is also worth noting that they were better than those obtained with 
sprocket and the pairwise versions of ConvS2S-VC and RNN-S2S, 
all of which were trained under a speaker-dependent closed-set condition.

We further evaluated the performance 
of the any-to-many model under an open-set condition 
where the speaker of the test utterances is unseen in the training data.
We used the utterances of the speaker lnh (female) as the test input speech.
The results are shown in Tab. \ref{tab:all_m2o_unseen}.
For comparison, 
Tab. \ref{tab:sprocket_lnh} shows results of 
sprocket performed on the same speaker pairs
under a speaker-dependent closed-set condition.
As these results show, the proposed model with the open-set any-to-many setting
still performed better than sprocket, even though sprocket had an advantage in
both the training and test conditions.

\begin{table}[t]
\caption{MCDs and LFCs obtained with
the real-time system settings.}
\centering
\begin{tabular}{l| l V{3} c|c}
\thline
\multicolumn{2}{c V{3}}{Speaker pair}&\multicolumn{2}{c}{Measures}\\\hline
source&target&MCD{\tiny (dB)}&LFC\\\thline
\multirow{3}{*}{clb} 
   &bdl&$6.77\pm .10$&$0.821$\\
   &slt&$5.95\pm .08$&$0.829$\\
   &rms&$6.35\pm .10$&$0.764$\\\hline
\multirow{3}{*}{bdl} 
   &clb&$5.99\pm .10$&$0.820$\\
   &slt&$6.32\pm .10$&$0.821$\\
   &rms&$6.54\pm .15$&$0.797$\\\hline
\multirow{3}{*}{slt} 
   &clb&$5.84\pm .10$&$0.818$\\
   &bdl&$6.92\pm .16$&$0.810$\\
   &rms&$6.48\pm .09$&$0.789$\\\hline
\multirow{3}{*}{rms} 
   &clb&$6.17\pm .11$&$0.789$\\
   &bdl&$6.74\pm .13$&$0.856$\\
   &slt&$6.42\pm .12$&$0.833$\\\hline
\multicolumn{2}{c V{3}}{All pairs}&$6.37\pm .04$&$0.812$\\\thline
\end{tabular}
\label{tab:all_causal}
\end{table}

\subsubsection{Performance with real-time system settings}

We evaluated the MCDs and LFCs obtained with
the many-to-many model under the real-time system 
setting described in \refsubsec{real-time}.
The results are shown in Tab. \ref{tab:all_causal}.
As the results show, the MCDs and LFCs 
were only slightly worse than those obtained with the default setting despite 
the disadvantage of using causal convolutions for all the networks 
and forcing attention matrices to be exactly diagonal (instead of having them be predicted).
A comparison of Tab. \ref{tab:all_causal}
with the results obtained with sprocket in 
Tabs. \ref{tab:mcd_baseline_comp} and \ref{tab:lfc_baseline_comp} 
may also show how well  
the proposed method can perform with the real-time system setting.

\begin{figure*}[t!]
\centering
\begin{minipage}[t]{.43\linewidth}
\centering
  \centerline{\includegraphics[width=.98\linewidth]{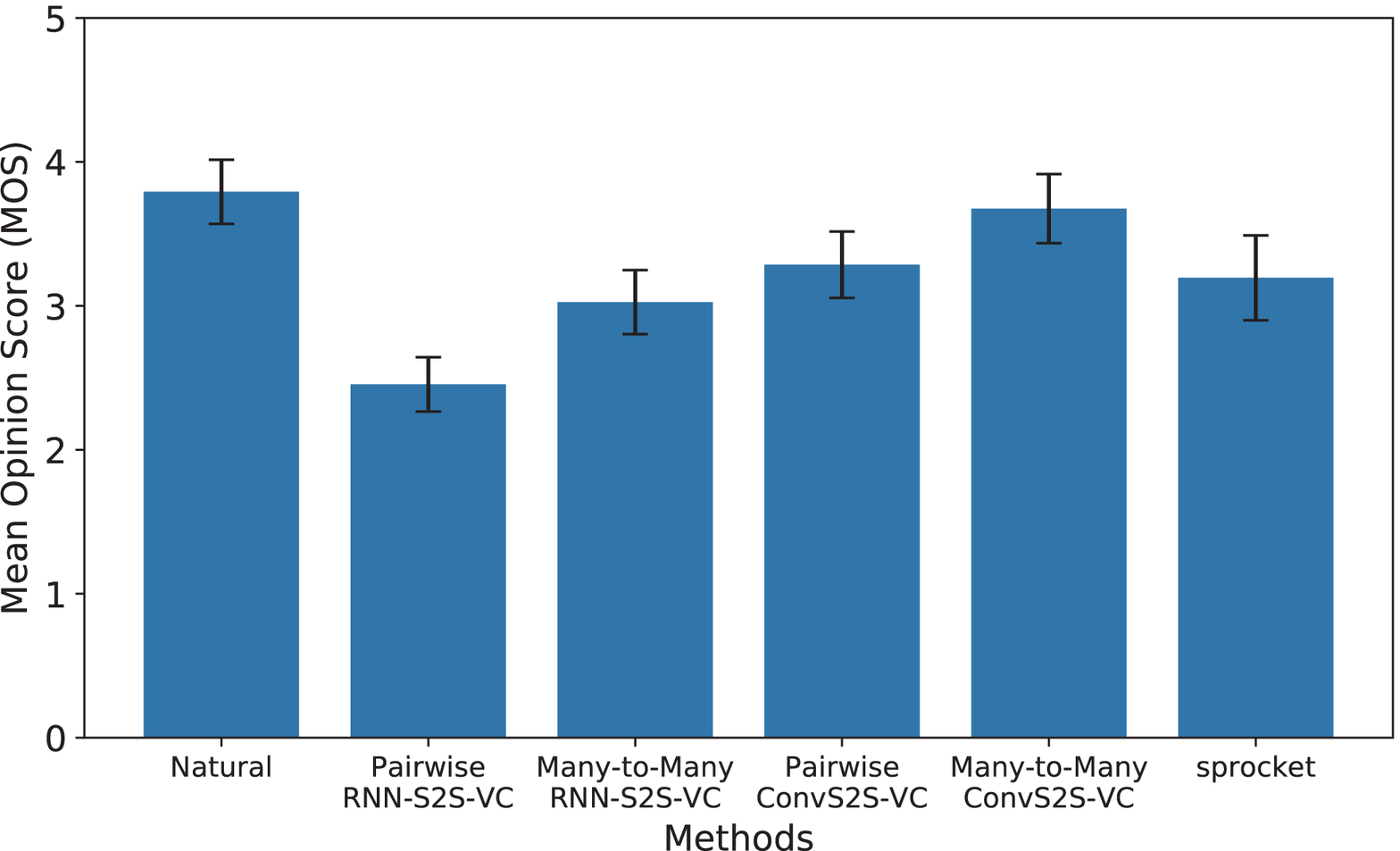}}
  \vspace{-1ex}
  \caption{Results of the MOS test for sound quality.}
  \label{fig:MOS_qlt}
\end{minipage}
\centering
\begin{minipage}[t]{.43\linewidth}
\centering
  \centerline{\includegraphics[width=.98\linewidth]{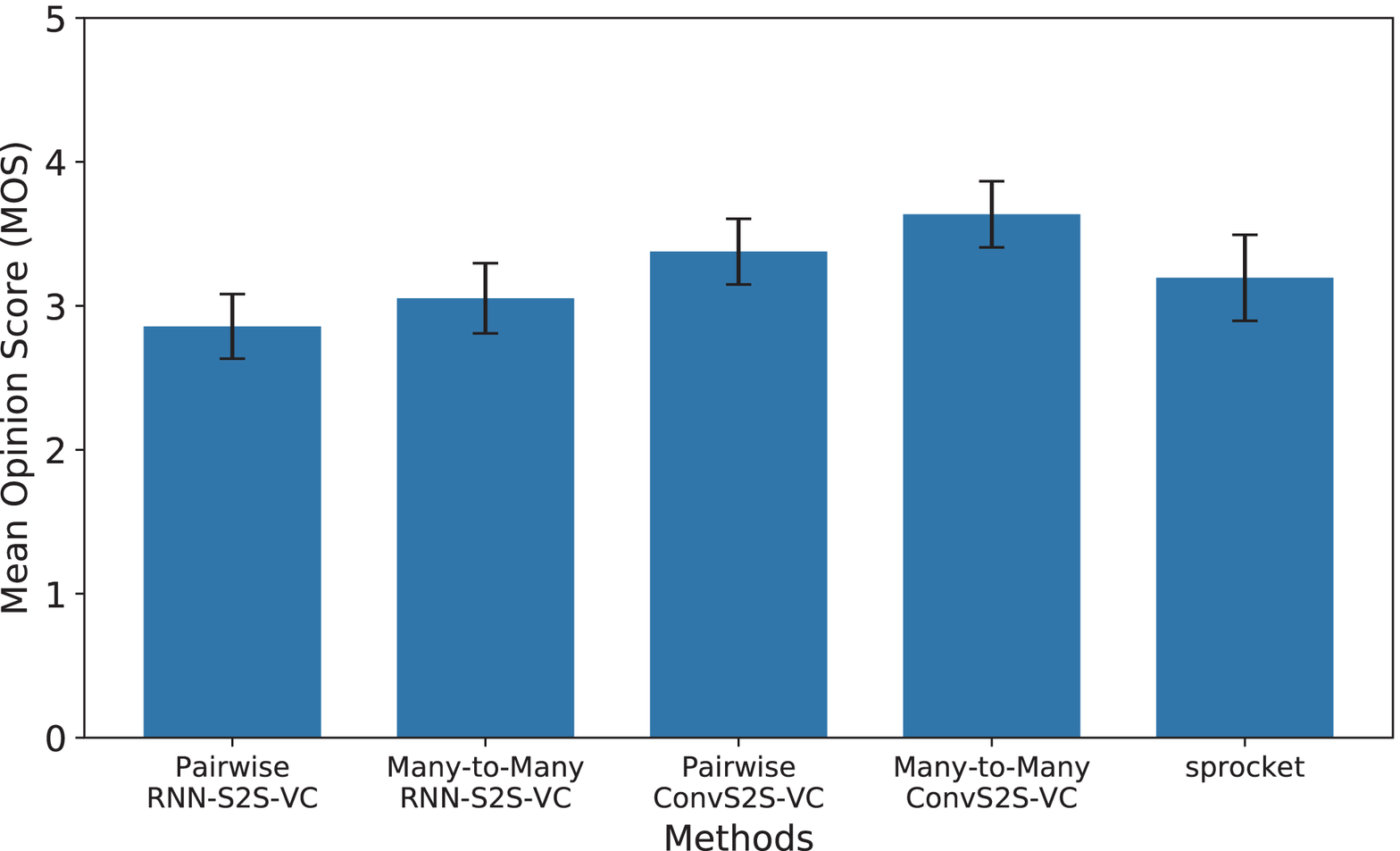}}
  \vspace{-1ex}
  \caption{Results of the MOS test for speaker similarity.}
  \label{fig:MOS_sim}
\end{minipage}
\end{figure*}

\subsection{Subjective Listening Tests}

We conducted mean opinion score (MOS) tests to compare the sound quality 
and speaker similarity 
of the converted speech samples
obtained with the proposed and baseline methods.

With the sound quality test, 
we included the speech samples 
synthesized in the same way as the proposed and baseline methods (namely the WORLD synthesizer) using
the acoustic features directly extracted from real speech samples. 
Hence, the scores of these samples are expected to show the upper limit of the performance. 
We also included speech samples produced using the pairwise and many-to-many versions of RNN-S2S-VC and sprocket in the stimuli.
Speech samples were presented in random orders to eliminate bias as regards the order of the stimuli. 
Ten listeners participated in our listening tests. 
Each listener was presented 6 $\times$ 10 utterances 
and asked to evaluate the naturalness
by selecting 5: Excellent, 4: Good, 3: Fair, 2: Poor, or 1: Bad for each utterance.
The results are shown in \reffig{MOS_qlt}.
As the results show, 
the pairwise ConvS2S-VC performed slightly better than sprocket and
significantly better than the two versions of RNN-S2S-VC.
The many-to-many ConvS2S-VC performed better than all other methods,
revealing the effect of the many-to-many extension, and reached close to 
the upper limit obtained with the analysis and synthesis technique.

In the speaker similarity test, 
each subject was given a converted speech sample and 
a real speech sample of the corresponding target speaker 
and was asked to evaluate how likely they are to be produced by the same speaker by selecting 
5: Definitely, 4: Likely, 3: Fair, 2: Not very likely, or 1: Unlikely.
We used converted speech samples generated by the pairwise and 
many-to-many versions of RNN-S2S-VC and sprocket for 
comparison as with the sound quality test. 
Each listener was presented 5 $\times$ 10 pairs of utterances.
As the results in \reffig{MOS_sim} show,
both the pairwise and many-to-many versions of ConvS2S-VC
performed better than all other methods.

\subsection{Audio examples of various conversion tasks}

Although we only considered a speaker identity conversion task 
in the above experiments, 
ConvS2S-VC can also be applied to other tasks. 
Audio samples of ConvS2S-VC tested on several tasks, 
including 
speaker identity conversion,
emotional expression conversion, 
electrolaryngeal speech enhancement, and
English accent conversion,
are provided at \cite{Kameoka2019url}.
From these examples,
we can expect that 
ConvS2S-VC can also perform reasonably well in 
various tasks other than speaker identity conversion.

\section{Conclusions}

This paper proposed a voice conversion (VC) method based on the ConvS2S learning framework.
The proposed method provides a natural way of converting 
the $F_0$ contour, speaking rate, and rhythm as well as the voice characteristics of input speech and 
the flexibility of handling many-to-many, 
any-to-many, and real-time VC tasks 
without relying on automatic speech recognition (ASR) models and text annotations. 
Through ablation studies, we demonstrated the individual effect of each of
the ideas introduced in the proposed method.
Objective and subjective evaluation experiments on a speaker identity conversion task
showed that the proposed method could perform better than baseline methods. 
Furthermore, 
audio examples showed 
the potential of 
the proposed method to perform well in 
various tasks including emotional expression conversion, 
electrolaryngeal speech enhancement, and English accent conversion.



\section*{Acknowledgments}

This work was supported by JSPS KAKENHI 17H01763 and JST CREST Grant Number JPMJCR19A3, Japan.

\ifCLASSOPTIONcaptionsoff
  \newpage
\fi



\bibliographystyle{IEEEtran}
\bibliography{Kameoka2018arXiv11}
%

%



\end{document}